\documentstyle[aps,preprint]{revtex}
\begin{document}
\draft
\title{Aharonov-Bohm Order Parameters for Non-Abelian Gauge Theories}
\author{Hoi-Kwong Lo}
\address{
 School of Natural Sciences, Institute for Advanced Study, Olden Lane,\\
 Princeton, NJ 08540, U.S.A.
}
\date{\today}
\preprint{IASSNS-HEP-95/8}
\maketitle
\mediumtext
\begin{abstract}
The Aharonov-Bohm effect
has been invoked to probe
the phase structure of a gauge theory. Yet in the case
of non-Abelian gauge theories, it proves difficult to formulate
a general procedure that unambiguously specifies the realization of the
gauge symmetry, e.g. the unbroken subgroup. In this paper, we propose a set
of order parameters that will do the job.
We articulate the fact that any useful Aharonov-Bohm experiment necessarily
proceeds in two stages: calibration and measurement.
World sheets
of virtual cosmic string loops can wrap around test charges, thus changing
their states relative to other charges in the universe.
Consequently, repeated flux
measurements with test charges will not necessarily agree.
This was the main stumbling block to
previous attempts to construct order parameters for non-Abelian
gauge theories. In those works, the particles that
one uses for calibration and subsequent measurement
are stored in {\em separate} ``boxes''.
By storing all test particles
in the {\em same} ``box'', we show how quantum fluctuations can be overcome.
The importance of gauge fixing
is also emphasized.
\end{abstract}
\medskip
\section{Introduction}
\label{sec:intro}
A gauge theory can have an interesting phase diagram. Depending on
it Higgs structure and on the parameters of the Higgs potential,
the theory may be in a Coulomb phase, a Higgs phase or a confinement
phase. Order parameters that distinguish among the various phases
have been proposed. Consider {\em pure} $SU(N)$
gauge theory. The Wilson loop operator
\begin{equation}
W(C) = tr \left[ P \exp \left( ig \oint_C a \cdot dx \right) \right]
\label{Wilson}
\end{equation}
may be regarded as an insertion of a classical source of
charges, transforming
as the defining representation of $SU(N)$, that propagates along the
worldline $C$. In the confinement phase, $C$ is the boundary of
the worldsheet of an electric flux tube. For large loops,
$W(C)$ exhibits area-law behavior
\begin{equation}
W(C) \sim \exp[ -\kappa  A(C) ],
\label{Areal}
\end{equation}
where $A(C)$ is the minimal area of
a surface bounded by $C$, and $\kappa$ is the string
tension.
In the Higgs phase, electric flux is screened, and the Wilson
loop has the perimeter-law behavior
\begin{equation}
W(C) \sim \exp[ -\mu P(C) ],
\label{Paral}
\end{equation}
where $P(C)$ is the length of $C$.
Thus, $W(C)$ is a useful order parameter for pure $SU(N)$ gauge theory.

Once dynamical quarks (a matter field that transforms non-trivially under
the center $Z_N$ of $SU(N)$) are introduced, the confining
and Higgs phases can no longer be distinguished by the above criterion.
Quark-antiquark pairs appear as quantum fluctuations, allowing electric
flux tube to break. The Wilson loop therefore
always obeys the perimeter law.
In the case of $SU(N)$ discussed above, it is widely believed that
no sharp boundary exists between the confinement
phase and the Higgs phase of interest \cite{tHooft}.

However,
two types of Higgs phases are possible, depending on
whether the center $Z_N$ is manifest or broken.
If the $Z_N$ symmetry is manifest, there will be a $Z_N$ superselection
rule. If it is unbroken, no such superselection rule exists.
It is presumably the Higgs phase with spontaneous broken
$Z_N$ that is indistinguishable from the confining phase.

How can we distinguish between the two types of Higgs phases?
Topological defects are potentially useful.
In this paper, we assume, for simplicity, that
a gauge theory with a (simply-connected)
symmetry group $K$ is spontaneously broken in two stages:
first to a discrete subgroup $G$ at mass scale $v_1$, then to
$H$, a subgroup of $G$ at mass scale $v_2 \ll v_1$.
We will focus on the second stage of the symmetry breakdown
and construct a set of order operators for our investigation.
The first stage of symmetry
breakdown gives rise to topological vortices (in $2+1$ dimensions)
and cosmic strings (in $3+1$ dimensions).
A cosmic string carries a ``flux'' which is labelled by an element of the
unbroken group $G$. [Vortices
are classified by the homotopy group $\pi_1 (K/G) $ \cite{Vilenk,Preski}.
It follows from
the exact homotopy sequence $\cdots \to
\pi_1(K) \to \pi_1 (K/G) \to \pi_0 (G) \to \pi_0 (K) $ that
 $\pi_1 (K/G) \simeq \pi_0(G) \simeq G$ under the assumptions that
$K$ is simply-connected and $G$ is discrete. To be more
precise, the spectrum of stable vortices only spans $G$ as a vortex
associated with an element $g$ of $G$ may be unstable to the decay
into two or more vortices of the same total flux. This is,
however, of no interest to us.]
Notice that there is a one-one correspondence between the elements
of the unbroken group and the topological classes of stable string loops.
This remains true even when $G$ is spontaneously broken into $H$,
in which case the topological classes of stable string loops are labelled by
elements of $H$.
(An element, $a$, of $G$ which is not in $H$ is not
associated with isolated string loops, but with string loops
that are bounded to domain walls.)
Therefore, by reading off the spectrum of stable
strings, we can determine the unbroken
group and decide whether the second stage of symmetry breakdown has
occurred.

So, the question becomes: how can we read off the fluxes of
the stable vortices in our theory?
A string generally has
long range Aharonov-Bohm \cite{AhaBoh}
type interactions with various particles:
the wave function of a particle acquires a non-Abelian
phase when it is covariantly transported around a string
\cite{KraWil,PreKra}.

This simple phenomenon has deep consequences.
Since the Aharonov-Bohm interaction is long range and no local
operator can destroy an object with an infinite-range interaction,
gauge theories with such interactions obey non-trivial
superselection rules. The structure of the superselection sector
can be invoked to distinguish among the various possible phases of
a gauge theory. Moreover, the Aharonov-Bohm interaction exposes the
limitations of the {\em classical} no-hair conjecture in black hole physics.
A black hole may carry quantum numbers that are detectable only
by means of {\em quantum mechanical}
interference experiments with cosmic strings \cite{CoPrWi}.
By sending particles around the various string loops and
measuring the non-Abelian Aharonov-Bohm factors
acquired by them, we can
read off the spectrum of stable strings. The
usage of the Aharonov-Bohm effect between charged particles
and cosmic strings to
probe the unbroken group of a non-Abelian gauge theory was proposed by
Preskill and Krauss \cite{PreKra}.
It requires a framework that takes full account of
the effect of virtual particles and virtual string loops.
Nevertheless, generalization to non-Abelian
gauge theories turned out to be very elusive. In spite of
much progress in our understanding of the subtler aspects of
non-Abelian gauge theories \cite{BuLoPr,Alford,BuLePr,Bavade,LoPres}
it proves difficult \cite{FreMar,AlfMar,qft} to formulate
a general procedure that unambiguously specifies the realization
of the gauge symmetry.

In this paper, we construct a set of order parameters that will
do the job, elaborating on our key results stated in Ref. \cite{orderp}.
These order parameters
are closely related
to some operators
investigated by Alford {\em et al.} three years ago \cite{qft}.
They, however, immediately rejected their operators
because, in their original forumlation, they were plagued
by quantum fluctuations.

As we will see below, any useful Aharonov-Bohm
experiment to determine the flux of a string loop
necessarily proceeds in two stages: calibration and subsequent measurement.
Both stages involve interference experiments with two beams
of charged particles
one of which traverses the string loop while the other just
sits at the base point.
To construct an order parameter for non-Abelian gauge theories,
the effects of virtual
string loops need to be considered.
The formulation by Alford {\em et al.} corresponded
to storing the charged particles
that one uses for calibration and those for measurement in separate
boxes. (See FIG. 1a.)
It was only because of this decoupling of the two that quantum
fluctuations due to virtual string loops may spoil the result.
What happen is that a small world
sheet of virtual
string loop can wind around the box which contains
all the calibrating particles, changing
their state relative to those we used for subsequent measurement
(that are stored in another box). As a result, the subsequent
flux measurement gives an answer different from the calibrated value.
If we keep both types of particles
in the same box (and for the subsequent measurement, an interference
experiment is performed between a wavepacket that is kept in the box
and another that is parallel transported around the string
of interest) (See FIG. 1b.), we argue that the effect of virtual strings can
be
safely ignored.

Another important issue that has been overlooked in
previous work \cite{AlfMar,qft} is gauge fixing.
Recall that we are interested in studying the symmetry breakdown
of $G$ into a subgroup $H$.
Without gauge fixing,
an element $h$ in an unbroken group $H$ has no invariant meaning.
Under a global gauge transformation by $g \in G$, $h \to ghg^{-1}$.
However, after symmetry breakdown
strings with fluxes $h$ and $ghg^{-1}$ are generally
{\em not} gauge invariant.
To test whether a symmetry breaking has occurred, one has to
choose a field $\phi$ as a candidate for the Higgs field,
gauge fix $\phi = \phi_0$ and consider $H( \phi_0)$ and its
conjugacy classes and representations.
By dealing with the issues of quantum fluctuations and
gauge fixing squarely, we see clearly how a gauge group $G$ is reduced to
an effective subgroup $H$ at low energies.
Subsequent symmetry
breaking of $H$ can be studied in a similar manner.

We emphasize that the idea of using the Aharonov-Bohm effect to
probe the phase diagram is rather general. In particular, the
existence of stable cosmic strings is {\it not} a necessary
condition. Nor is the symmetry group
required to be discrete. (After all, the Aharonov-Bohm effect exists in
ordinary
QED which has a continuous symmetry.)
One can always imagine
setting up localized magnetic flux tubes and studying their long
ranged Aharonov-Bohm
interactions with charged particles. Suppose symmetry breaking
now occurs. Some of those flux tubes will then become the boundaries of
unstable domain walls and the Aharonov-Bohm interactions will be
modified in such a way to be consistent with the symmetry of the unbroken
group.
It is only for simplicity that
we restrict our discussion to gauge theories with discrete symmetry only.
Besides, we sometimes make use of non-gauge
invariant operators. The reader should, however, bear in mind
that the whole discussion can be recast in
objects that are
gauge-invariant with respect to the unbroken low energy symmetry group.

The organization of the paper is as follows. Section 2 concerns
the relation between Dirac quantization condition and the Aharonov-Bohm
effect. In Section 3, we discuss the basic ideas behind the construction
of the Aharonov-Bohm order parameters.
How quantum fluctuations can change the state of test particles
and affect the result of a subsequent Aharonov-Bohm experiment
is the main subject of our study in Section 4.
Sections 5 and 6 concern respectively quantum fluctuations
and gauge fixing, two crucial issues in the construction.
Finally, we present our
discussions and conclusions in Section 7. In the Appendix,
we review the application of the order parameters to $Z_2$ gauge-Higgs
system by Preskill and Krauss.
\section{Dirac Quantization Condition and
the Aharonov-Bohm effect}
\label{sec:Dirac}
In his seminal paper of 1931, Dirac \cite{Dirac}
proposed that the quantization
of electric charge can be ``explained'' by postulating the existence
of isolated magnetic poles. Specifically, he showed that,
for a charged particle of charge $e$ moving in the field of a
magnetic monopole of strength $\mu_0$,
the consistency of its quantum mechanics demands that the product
of the two charges satifies
$e \mu_0 ={1\over 2} n \hbar c. $ This is the well-known Dirac quantization
condition. Note that it has  an explicit dependence on
Planck's constant, and therefore on the quantum theory. Moreover,
it shows a perfect symmetry between electricity and magnetism.

Generalization of this simple condition has deep implications.
For instance, consider an underlying gauge theory with gauge group $G$ which
is spontaneously broken into a subgroup $ SU(N)/Z_N$. This
theory admits magnetic monopoles with $Z_N$ magnetic charges.
Fields
that transform non-trivially under $Z_N$ and the monopole with
minimal magnetic charge taken together
do not satisfy the Dirac condition. One immediately sees that
the quantum mechanics of a system of a
free minimal monopole and a free $Z_N$ charge
is inconsistent. Thus, one is led to the conclusion that
either the minimal monopole or the $Z_N$ charge must be confined.
This is the basic concept behind t' Hooft's discussion on quark
confinement \cite{tHooft}.
The t' Hooft  operator $B_n(C')$ essentially introduces, as a
classical source, a $Z_N$ monopole worldline along the curve $C'$.
It is the magnetic analogue of
the Wilson loop operator $A^{\nu}(C)$, which introduces, as
a classical source, a worldline of a charged particle in the
representation $\nu$. t' Hooft considered a
Green's function $\langle B_n(C') A^{\nu}(C) \rangle$. If the Dirac
quantization condition is not satisfied, this object is generally
multivalued.

If the charged particle is regarded as free, it
will see the Dirac string associated with the monopole. That is
to say that the Dirac string is physical and it has a long-range
Aharonov-Bohm interaction with the charged particle. The
Green's function can be made single-valued if a world sheet $\Sigma^*$ of
the Dirac string is chosen. i.e.
$\langle B_n(C',\Sigma^*) A^{\nu}(C) \rangle$ is single-valued. If $C'$
shrinks to a point, we replace the operator $B_n(C',\Sigma^*)$ by
$F(\Sigma^*)$ which introduces a closed world sheet of string.
We emphasize that the Aharonov-Bohm effect is quantum mechanical.
It can be determined only through interference experiments.
(Incidentally, the Aharonov-Bohm interactions between magnetic
monopole and electric flux tubes can be discussed in a totally
analogous manner \cite{qft}: If a monopole is regarded as free, (i.e. there is
no physical Dirac string) charged particles which do not satisfy the Dirac
quantization condition are confined. Therefore, it is appropriate
to consider operators $G^{\nu}(\Sigma)$ and  $B_n$
which respectively introduce a world sheet of an
electric flux tube and a worldline of a
magnetic monopole.)
As we will see in next section, the operators $F$ and $A$ will
play a key role in the construction of order parameters for
gauge theories.

\section{Order Parameters}
In this section, we consider the order parameters
for Abelian gauge theories proposed by Preskill and Krauss \cite{PreKra}
and elaborated by Alford {\em et al.} \cite{qft}.
By the well-known
Elitzur's theorem \cite{Elitzur}, the Higgs field $\phi$ is not a
true order parameter because it has no gauge invariant meaning.
The order parameters proposed by Preskill and Krauss make
essential use of the Aharonov-Bohm interactions between
cosmic strings and charged particles.

While our ideas are general, for simplicity, we shall discuss
the concepts for only the case of finite gauge groups\cite{PreKra,AlfMar,qft}.
More specifically, we consider
a discrete gauge group $G$ which arises as a result of a symmetry
breaking of a simply-connected group $K$. We are interested in
investigating the further symmetry breakdown of $G$ into
a subgroup $H$ at low energies.

The symmetry breakdown of $K$ into
$G$ leads to the existence of stable vortices labelled by elements
of $G$. (More details will be discussed in the next Section.)
Suppose we prepare (and calibrate) a set of cosmic strings.
In a free $G$ charge phase, a particle scattering off a cosmic string
will recover the flux of the string as a non-Abelian Aharonov-Bohm factor.
If $G$ is broken into $H$ instead,
then the elements of $G$ that are not in $H$ are not associated
with isolated cosmic strings, but with strings that are boundaries of domain
walls.
Such domain walls are
unstable and will decay via spontaneous nucleation
of string loops \cite{Vilenk,Preski}. Consider
a string of flux $a \notin H$ bounded by a domain wall.
Many holes eventually appear in the wall bounded to $a$. They collide
with one another. Ultimately, the one with the least string tension,
$b$, will dominate the decay. As a result of this decay, the $a$ string
is turned into a composite string with total flux $ab^{-1} \in H$.
Hence, a particle scattering off a cosmic string in a free $H$ charge
phase will only acquire phases that are associated with the elements of $H$.
[We must include the possibility that the set of strings with
the least
string tension has more than one elements. For instance, a string with the
least string tension can have a flux $b$ or $b'$.
Therefore, the Aharonov-Bohm
experiment may show that the composite flux after the decay of the
domain wall is $ab$ or $ab'$.]

The basic idea of the Aharonov-Bohm order parameters is the
following. Since, in the case of discrete gauge theories,
there is always a one-one correspondence between the topological classes
of stable vortices and the elements of the unbroken group
(see the Introduction and the next Section for details), we can
figure out
the manifest symmetry group just by reading out the
spectrum of stable vortices. To read off the spectrum, we use the
Aharonov-Bohm effect. Physically, we
proceed in two stages: 1) Prepare (Calibrate)
a {\it localized} vortex for each element of $G$. 2)
Send particles around those vortices to measure the associated Aharonov-Bohm
factors.
The results of the measurements can tell
us whether the symmetry breaking from $G$ to $H$ has
occurred or not.

Let us consider the first stage of an Aharonov-Bohm experiment:
Calibration of string flux\cite{AlCoMa}.
We need an operator which inserts,
as a classical source, a string world sheet
of flux $a$. It was suggested in \cite{KraWil} that when a $U(1)$
gauge symmetry
is spontaneously broken into $Z_N$, the discrete $Z_N$ charge
$Q_{\Sigma^*}$ contained in a closed surface
$\Sigma^*$ can still be measured via
the Gauss law:
\begin{equation}
F(\Sigma^*)= \exp \left({2 \pi i \over N} Q_{\Sigma^*} \right)
=\exp  \left( {2 \pi i \over
Ne} \int_{\Sigma^*} E \cdot ds \right). \label{gauss}
\end{equation}
[$F(\Sigma^*)$ is closely related to the 't Hooft loop operator \cite{tHooft}.
Strictly speaking, Eq.\ (\ref{gauss}) is incorrect as an operator
statement. Charge can be defined in two different ways: the Coulomb charge and
the Aharonov-Bohm charge. The Coulomb charge is the one
that enters in the Coulomb force between two charged particles. It is
screened in a Higgs phase. Therefore, the Gauss law actually
gives zero for the Coulomb charge. However, the Aharonov-Bohm interactions
between confined flux tubes and charges are
unscreened (See, for example, \cite{cherns}). In Eq.\ (\ref{gauss}),
we are just making the heuristic statement that charges can be
detected by their Aharonov-Bohm interactions with vortices.]

Now we turn to the operator which introduces classical charges into the
system. An obvious choice would be the Wilson loop
operator $W^{\nu}(C)$ where $\nu$ is an irreducible representation of
the gauge group, $G$.
Therefore, one might naively expect
$F(\Sigma^*) W^{\nu}(C)$ to be the order parameter.
This is not quite correct
because quantum mechanical fluctuations near
the surface
$\Sigma^* $ cause an area law decay of the {\em modulus\/} of
$F(\Sigma^*) \sim \exp \left(- \kappa A({\Sigma^*}) \right)$. Fortunately,
the {\em phase\/} of $F(\Sigma^*)$ remains unscreened and we can isolate
it by dividing out its vacuum expectation value and obtain
$F(\Sigma^*) \over \langle F(\Sigma^*) \rangle $\cite{PreKra} .
Similarly, quantum fluctuations also
lead to the exponential decay of the expectation value of $W(C)$.
Therefore, the true order parameter for Abelian gauge theories is
\cite{PreKra}
\begin{equation}
A^{\nu}(\Sigma^{*}, C) = {F(\Sigma^{*})
W^{\nu}(C) \over  \langle F(\Sigma^*) \rangle
\langle W^{\nu}(C)\rangle}.
  \label{abelian}
\end{equation}
In the free $Z_N$ charge phase, the order parameter (for the fundamental
representation) gives
\begin{equation}
\lim \langle A(\Sigma^{*}, C) \rangle
 = \exp \left({2 \pi i \over N} k( \Sigma^* ,C) \right) .
\label{free}
\end{equation}
Here the
order parameter takes a value that is independent
of the details of $\Sigma^*$ and $C$ so long as
the limit is taken with $\Sigma^* $ and $C$ increasing to infinite
size, and with the closest approach of $\Sigma^*$ to $C$ also approaching
infinity; $ k( \Sigma^* ,C)$ denotes the linking number of the surface
$\Sigma^*$ and the loop $C$.
($\Sigma^* $ and $C$ have to be far away from each other because we
are only interested in the {\it long range} Aharonov-Bohm interactions.)
On the other hand, if there are no free
$Z_N$ charges, then we have
\begin{equation}
\lim \langle A(\Sigma^{*}, C) \rangle
 = 1 .
\label{nofree}
\end{equation}
The non-analytical behavior of $ A(\Sigma^{*}, C)$ guarantees that the
two phases are separated by a well-defined phase boundary.
An order parameter can be easily generalized to probe the realisation of
any Abelian discrete gauge symmetry.

\section{Quantum fluctuations}
\label{sec:Hamil}
The case of non-Abelian gauge theories is more subtle.
In this Section, we will show how quantum fluctuations
can affect the result of a non-Abelian Aharonov-Bohm experiment,
thus making the construction of the order parameter much
more difficult.

Consider, in two spatial dimensions,
a simply-connected
gauge group $K$ which gets spontaneously broken into a
discrete, non-Abelian subgroup
$G$. This pattern of symmetry breaking will admit stable classical
vortex solutions. Since the size of the vortex core is of the
order of the inverse symmetry breaking scale, it is
almost pointlike at low energies.
We shall, therefore, ignore the core of the vortex and idealize
it as a point singularity.
The ``flux'' carried by a vortex is labelled
by an element of the unbroken group $G$. To assign a group
element to the vortex, we {\em arbitrarily} choose a ``base point''
$x_0$ and a path $C$, beginning and ending at $x_0$, that winds around
the vortex. The effect of parallel transport in the
gauge potential of the vortex is then encoded in
the untraced Wilson loop operator \cite{Bais,Bucher}
\begin{equation}
U^{\nu} (C, x_0) = P \exp \left( i \int_{C,x_0} A \right)
\label{untraced1}
\end{equation}
where $P$ denotes path ordering.
Suppose a particle is in the initial state
$| u \rangle$. When it
winds around
a vortex (or a cosmic string loop in three dimensions), its
state becomes
$U^{\nu} (C, x_0) | u \rangle$.
The matrix $U^{\nu} (C, x_0) $
specifies a group element in the subgroup $G(x_0)$ of
$K$ that preserves the Higgs condensate at the point $x_0$, since
transport of the condensate around the vortex must return it to its
original value. If $G$ is discrete, then the element assigned
will remain unchanged as the path $C$ is smoothly deformed, as long
as the path never crosses the cores of any vortices. (The gauge
connection is locally flat outside the vortex cores, with curvature
singularities at the cores.)

In the non-Abelian case, the flux, $a$, of a vortex
is {\em not\/} a gauge invariant quantity. Upon
a global gauge transformation by $g$,
$a \to g a g^{-1}$. One might naively identify two vortices in
the same conjugacy class as identical. However, this is not
quite correct because there is only one
overall global gauge degree of freedom. For example, if there are two
vortices of flux $a$ and $b$, upon a global gauge transformation by $g$, we
have
$a \to gag^{-1}$, $b \to g b g^{-1}$ \cite{PreKra,LoPres,Bucher,WilcWu}.
It is important to note that,
two vortices with conjugate but different fluxes (in some gauge)
are not
identical to each other. Consider two vortices of conjugate
but different fluxes, $a$ and $b=cac^{-1}$ in some gauge. They are clearly
different in this gauge. (As we will see
below, a particle that winds around the $a$ vortex first
and then in the inverse sense around the $b=cac^{-1}$ vortex will pick up the
Aharonov-Bohm factor depending on $ca^{-1}c^{-1}a$. Here our convention
for multiplication is from right to left. If these two vortices
had the same flux, the Aharonov-Bohm factor would be trivial instead.)
Now under a global gauge transformation,
by $g$, $a \to gag^{-1}$ and $cac^{-1} \to gcac^{-1}g^{-1}$
and the two fluxes clearly remain different. Notice also
that the one-one correspondence
mentioned in the last Section is between the spectrum of stable
vortices and the {\it elements} of the unbroken group rather
than its conjugacy classes. It is, therefore, crucial
to be able to distinguish between vortices that are
associated with different elements in the same conjugacy
class. To obtain this one-one correspondence, one has to fix
a gauge. Fortunately, a change of gauge merely amounts to
a change of basis and the one-one correspondence still exists in the
new gauge. The analogy with isospin may be helpful for
understanding why vortices of conjugate but
different fluxes should be regarded as non-identical. In an isospin symmetric
universe, it is a matter of convention to call an object
a neutron rather than a proton. However, once we call an object
a neutron, our convention has been fixed  and we will be able to
distinguish a proton from a neutron by comparing their isospins.

If a particle that transforms as
an irreducible representation $(\nu)$ traverses a string of flux $a$, the
Aharonov-Bohm phase that it acquires, when averaged over a basis
for the representation, is
\begin{equation}
{1 \over n_{\nu}} \chi^{\nu} (a) ,
\label{aphase}
\end{equation}
where $\chi^{\nu}$ and $n_{\nu}$ are the character and the dimension
of $({\nu})$. Now if two strings of fluxes $a$ and $b$ are patched
{\it incoherently}, the Aharonov-Bohm phase acquired by the particle
that travels around the two strings in succession is
\begin{equation}
{1 \over n_{\nu}} \chi^{\nu} (a){1 \over n_{\nu}} \chi^{\nu} (b)  .
\label{abphasei}
\end{equation}
If the two strings are combined {\it coherently} instead, the phase acquired
is
\begin{equation}
{1 \over n_{\nu}} \chi^{\nu} (ab) .
\label{abphasec}
\end{equation}
Thus, the Aharonov-Bohm factor associated with a coherent pair
of string is {\it not} just the product of the Aharonov-Bohm factors
associated with the two individual strings. This coherence property is the
hallmark of the non-Abelian Aharonov-Bohm effect.
%
%
%
%

The above discussion ignores the effect of quantum fluctuations.
Quantum fluctuations can spoil the coherence between
various strings. It is instructive to consider a double slit
experiment with {\it two} vortices of fluxes $g_1$ and $g_2$
placed between the two slits and using a particle-antiparticle pair
in the representation $(\nu)$. Let us put
the $g_2$ vortex in {\it front} of the two slits
and the $g_1$ vortex behind. Suppose the particle-antiparticle
pair is initially in the zero-charged state, i.e.,
\begin{equation}
|0, \nu \rangle =
{1 \over \sqrt n_{\nu}} \sum_i |e^{\nu}_i \otimes e^{* \nu}_i \rangle .
\label{zeroch}
\end{equation}

Let us first consider the case without quantum fluctuations.
When the particle traverses the double slit, the state
of the entire state will transform in the following manner:
\begin{equation}
|g_1,g_2 \rangle \otimes
{1 \over \sqrt n_{\nu}} \sum_i |e^{\nu}_i \otimes e^{* \nu}_i \rangle
\rightarrow
|g_1,g_2 \rangle \otimes
{1 \over \sqrt n_{\nu}} \sum_{ij}
 | e^{\nu}_j \otimes e^{* \nu}_i \rangle D^{(\nu)}_{ji} (g_1 g_2),
\label{transform1}
\end{equation}
where $D^{(\nu)}_{ji}$ are the matrix elements of the representation
$(\nu)$.

For the particle-antiparticle pair to annihilate, we project
onto the zero-charged state using the projection operator
\begin{equation}
P_{particle} = |0, \nu \rangle \langle 0, \nu |
\label{particle}
\end{equation}
to obtain
\begin{eqnarray}
& &|g_1,g_2 \rangle \otimes
{1 \over  \sqrt n_{\nu}}
 \sum_k |e^{\nu}_k \otimes e^{* \nu}_k \rangle \sum_{ijk'}
 \langle e^{\nu}_{k'} \otimes e^{* \nu}_{k'} |
 e^{\nu}_j \otimes e^{* \nu}_i \rangle
 {1 \over  n_{\nu}}
D^{(\nu)}_{ji} (g_1 g_2) \nonumber \\
&\rightarrow& |g_1,g_2 \rangle \otimes
|0, \nu \rangle {1 \over n_{\nu}} \chi^{\nu}(g_1 g_2) .
\end{eqnarray}
In other words, in the absence of virtual
processes, the interference pattern will determine the Wilson loop
to be ${1 \over n_{\nu}} \chi^{\nu}(g_1 g_2)$.

Let us now turn to quantum fluctuations. Notice that
two beams of particles are split and recombined in
an Aharonov-Bohm experiment.
For those
quantum fluctuations (such as virtual vortices
that wind around only one of the two beams)
whose effects do not depend on
the flux of vortex with which we are performing the
Aharonov-Bohm interference experiment, their effect
can simply be factored out. (Cf. Eq.\ (\ref{abelian}).) However, there
are quantum fluctuations (that affect both beams)
whose effects cannot be factored out:
consider the
spontaneous nucleation of
a charge-zero virtual vortex-antivortex pair in the
conjugacy class $[g']$, i.e.,
\begin{equation}
|0, [g'] \rangle=
{1 \over \sqrt n_{[g']}} \sum_{g \in [g']} | g, g^{-1} \rangle ,
\label{vortex}
\end{equation}
(here $n_{[g']}$ is the number of elements in the conjugacy class
$[g']$) in the region between the two slits and
the $g_1$ vortex. Suppose the vortex and antivortex
move apart just before the double slit experiment.
When a particle traverses the two slits,
the state of the entire system will change as follows:
\begin{eqnarray}
& &|g_1,g_2 \rangle \otimes
{1 \over \sqrt n_{[g']}} \sum_{g \in [g']} | g, g^{-1} \rangle \otimes
{1 \over \sqrt n_{\nu} }\sum_i |e^{\nu}_i \otimes e^{* \nu}_i \rangle
\nonumber \\
&\rightarrow &
|g_1,g_2 \rangle \otimes
{1 \over \sqrt n_{[g']}} {1 \over \sqrt n_{\nu}}
\sum_{g \in [g']} \sum_{ij} | g, g^{-1} ,
 e^{\nu}_j \otimes e^{* \nu}_i \rangle D^{(\nu)}_{ji} (gg_1g^{-1} g_2).
\end{eqnarray}
(i.e., for each $g \in [g']$,
the particle ``sees'' a flux $gg_1g^{-1} g_2$ rather than $g_1 g_2$.)
Now to make sure that the virtual vortex-antivortex pair
will annihilate, we apply the projection operator
\begin{equation}
P_{vortex}= |0, [g'] \rangle \langle 0, [g']| .
\label{provor}
\end{equation}
We also project onto the zero-charged state for the particle-antiparticle
pair by using the
operator
$P_{particle}$ defined in Eq.\ (\ref{particle}).
Therefore, we obtain
\begin{eqnarray}
& &|g_1,g_2 \rangle \otimes
{1 \over \sqrt n_{[g']}} {1 \over \sqrt n_{\nu}}
\sum_{g \in [g']} \sum_{ij} | g, g^{-1} ,
 e^{\nu}_j \otimes e^{* \nu}_i \rangle D^{(\nu)}_{ji} (gg_1g^{-1} g_2)
\nonumber \\
&\stackrel{P_{vortex} P_{particle}}{\longrightarrow} &
|g_1,g_2 \rangle \otimes |0, [g'] \rangle |0, \nu \rangle
 {1 \over n_{[g']}} \sum_{g \in [g']} \chi^{\nu}(gg_1g^{-1}g_2).
\end{eqnarray}
Notice that the Aharonov-Bohm factor acquired by the particle,
\begin{equation}
{1 \over n_{[g']}} \sum_{g \in [g']} \chi^{\nu}(gg_1g^{-1}g_2),
\label{pha}
\end{equation}
is generally different from the corresponding Aharonov-Bohm factor
without the virtual vortex pair (${1 \over n_{\nu}}\chi^{\nu} (g_1 g_2)$).
This result shows clearly that quantum fluctuations due to
virtual string loops can spoil the hallmark of the non-Abelian
Aharonov-Bohm effect---the coherence
between two strings in a double slit measurement\cite{private}.
This tends to create great difficulty in interpreting
the outcome of a non-Abelian Aharonov-Bohm experiment.
We will come
back to discuss how this difficulty can be overcome in Section 5.

A somewhat simpler but less precise way of stating our
result is that for non-Abelian gauge theories
repeated Aharonov-Bohm flux measurements
do not necessarily agree\cite{qft}.
Suppose we send out some of the charged particles to
calibrate a vortex and keep the rest elsewhere for later use.
A virtual vortex-antivortex
pair is spontaneously nucleated and the virtual vortex winds around
the charged particles that we use for calibration.
We now send out the remaining charged particles to measure the flux of
the vortex again.
The claim is that we will
find that its flux has been conjugated. As shown in
the spacetime diagram (FIG. 2),
the virtual vortex worldline has a non-trivial linking number with
the union of the following three objects: the
worldline of the vortex under consideration and the
worldlines of the charged particles that we use for the calibration
AND that of the ones that
we use for subsequent measurement.
The topological linking number shows that the virtual
vortex worldline conjugates the result
of the subsequent measurement relative to the calibrated value\cite{private}.

We remark that there are other kinds of quantum fluctuations in which
quantum numbers are exchanged between two objects. For instance,
an $i$ vortex can momentarily emit a $-1$ vortex, turning itself
into a $-i$ vortex. If this $-1$ vortex is absorbed by the
$j$ vortex, quantum numbers will be exchanged between the $i$ and
$j$ vortices. However, these types of quantum
fluctuations are uninteresting for our purposes and will
not be considered any further in this paper.

\section{Order Parameters for non-Abelian gauge theories}
\subsection{General Formulation}
We have seen in Section 3 how the Aharonov-Bohm effect
can be used to probe the phase diagram of an Abelian gauge theory.
In this section, we would like to extend the construction to non-Abelian
gauge theories.
This generalization turns out to be difficult and all
previous attempts have not been entirely successful \cite{AlfMar,qft} .
For ease of discussion, let us fix the gauge completely.
As discussed in Section 4, we shall regard vortices of
conjugate but different fluxes to be non-identical.
For each element $a \in G$, we need to define an operator $F_a(\Sigma^*)$
which introduces a world sheet $\Sigma^* $ of a string of flux $a$.
(The fact that the flux, $a$,
is not a gauge invariant quantity is not a problem because, at the end
of our discussion, we will apply the overall global gauge degree of freedom
to obtain gauge-invariant operators.)

For simplicity, consider the case of 2+1 dimensions.
How can we specify the worldline $\Sigma^*$
of a vortex of flux $a$ using the operator $F_a(\Sigma^*)$? One can imagine
assembling a laboratory of test particles at some
arbitrary base point $x_0$ and choosing
a basis for various representations there.
We then send two beams of particles to pass on either side of
the vortex, recombine the beams and study the resultant interference
pattern. In fact, a {\it sequence} of the Aharonov-Bohm experiments
has to be performed over time to specify the whole worldline of the vortex.
To localized the vortex worldline to (be close to) $\Sigma^*$, those
calibration experiments have to be done near to the vortex.
(To be more precise, for each of those calibration experiments, the
two beams involved are split only in a small region near $\Sigma^*$. One
of the beam then traverses the vortex and the two beams
are recombined immediately afterwards.)
The case of 3+1 dimensions is the same except that
now we have to specify the {\it worldsheet} of a string.

In a cubic four-dimensional
lattice formulation, it is convenient to put a string world sheet
on a closed
surface $\Sigma^* $ on the dual lattice. Let $\Sigma$ be the set of
plaquettes threaded by $\Sigma^*$.
Here comes the crucial point.
We pick our calibration paths:
for each plaquette $P$
in $\Sigma$, we choose a path, $l_P$, that runs from the base point $x_0$ to a
corner of the
plaquette \cite{qft}. Calibration of the plaquette is done
along the path $l_P P l_P^{-1} $.
The operator $F_a(\Sigma^{*}, x_0, \{l_P\})$ inserts, as a classical
source, a string world sheet $\Sigma^*$ calibrated
along paths $ \{l_P\}$ and
modifies the gauge
action in the following manner.
Suppose that the Euclidean plaquette action is
\begin{equation}
   S^{(R)}_{gauge, P}= - \beta \chi^{(R)} (U_P) + c.c. \label{oldgauge}
\end{equation}
   where
$R$ is some representation of the gauge group that defines the theory
and
$U_P = \prod_{l \in P} U_l $ associates with each plaquette (labelled by
P) the ordered product of the four $U_l$'s associated with the oriented
links of the plaquette.
The insertion of $F_a(\Sigma^{*}, x_0, \{l_P\})$ modifies the action
on each plaquette in $\Sigma$ to
\begin{equation}
 S^{(R)}_{gauge, P} \to
 - \beta \chi^{(R)} (V_{l_P}a V_{l_P}^{-1} U_P) + c.c. \label{newgauge}
\end{equation}
where
\begin{equation}
V_{l_P} = \prod_{l \in l_P} U_l.  \label{path}
\end{equation}
(This procedure can be generalized to insert {\em coherently\/}
many string loops using an operator $F_{a_1, a_2, \cdots, a_n}(\Sigma^{*}_1,
\Sigma^{*}_2, \cdots, \Sigma^{*}_n ,x_0, \{l_P\})$. See below. Note that
for coherent insertion, it is crucial to choose the same base point $x_0$
for all string loops. This
operator is not gauge invariant. Upon gauge transformation
by $g$ at the base point, it changes to  $F_{ga_1 g^{-1}, g a_2 g^{-1} ,
\cdots, g a_n g^{-1}}(\Sigma^{*}_1,
\Sigma^{*}_2, \cdots, \Sigma^{*}_n ,x_0, \{l_P\})$.)

Up to now, we have been vague about the choice of the
``tails'' $\{l_P\}$ (i.e. the paths for the calibration).
As it turns out, the choice is actually quite important. Unless the tails
$\{l_P\}$
are chosen in a judicious manner, because of quantum fluctuations,
there is no guarantee that
a subsequent measurement of the flux of a calibrated string
will recover the same result.

Now we turn to the operator which introduces classical charges into the
system. After gauge
fixing, all information of the non-Abelian Aharonov-Bohm effect is encoded
in
the {\em untraced\/} Wilson loop operator
\begin{equation}
U^{(\nu)}(C, x_0)= D^{(\nu)} \left( \prod_{l \in C} U_l \right)
\label{untrace}
\end{equation}
  where $C$ is a closed loop around $x_0$ and $\nu$
is an irreducible representation of the gauge group $G$. Conceptually,
after gauge fixing
all the matrix elements in
$U^{(\nu)}(C, x_0)$ can, in principle, be determined by interfering
charged particles in the
representation $\nu$
that traverse $C$ with those that stay at the base point \cite{AlCoMa}.
Just like
$F_a$, the operator $U^{(\nu)}(C, x_0)$ is not
gauge invariant.

When $\beta \gg 1$, the plaquettes are hard to excite. Therefore,
configurations with only a small number of frustrated plaquettes will
be important and we can expand in powers of $\exp(- \beta)$ (or equivalently
in terms of world sheets of virtual string loops).
When $F_a(\Sigma^{*}, x_0, \{l_P\})$ is inserted,
the configuration at weak couplings that has none of
its plaquette frustrated is an ``a-forest'' \cite{AlfMar}
in which, roughly speaking,
all the links
that intersect the minimal surface with boundary $\Sigma^*$
are of flux $a$. (FIG. 3.) Suppose the Wilson loop
links once with the string loop. Exactly one of its links, $l$
(say with a flux $V_l$), is an
$a$-forest link and this gives a flux $a$ for the Wilson loop.
More generally, in a phase with
free $G$ charges, and
in the leading order of
weak coupling perturbation theory,
the operator \cite{qft}
\begin{eqnarray}
\langle A^{\nu}_{a}
( \Sigma^{*} ,x_0,\{l_P\};C) \rangle
&=&{ \langle F_a(\Sigma^{*}, x_0, \{l_P\}) U^{(\nu)}(C, x_0) \rangle \over
\langle F_a(\Sigma^{*}, x_0, \{l_P\}) \rangle
\langle tr U^{(\nu)}(C, x_0)}\rangle \nonumber \\
&=& {1 \over n_{\nu}} D^{\nu} \left( a^{k (\Sigma^{*}, C)} \right)
\label{order}
\end{eqnarray}
 where ${k (\Sigma^{*}, C)}$ is the linking number
of the surface $\Sigma^{*}$ and the loop $C$ and the limit that
$\Sigma^{*}$ and $C$ are infinitely large and far away is taken.
This shows that, once a string loop is calibrated to be of flux $a$
along the paths $\{l_P\}$, a subsequent interference experiment
with a charged
particle will recover the same
non-Abelian Aharonov-Bohm factor.

\subsection{Quantum Fluctuations}
However, owing to quantum fluctuations,
higher order terms in the weak coupling expansion may spoil
this result \cite{qft}.
Recall that the dominant contribution in a weak coupling expansion
comes from configurations with a low density of frustrated
plaquettes (or equivalently, a low density of virtual string loops).
Now
in the definition of $ F_a(\Sigma^{*}, x_0, \{l_P\})$, for each plaquette
$P$, there is a long tail of links $l_P$ that connects it to $x_0$.
This is the calibration path for that particular plaquette.
Alford {\em et al.} have considered the choice
in which the long tails, $\{l_P\}$, from all the
plaquettes finally merge together and connect to the base point
through a single link which is not on the Wilson loop.
Essentially, they keep all
the charged particles for the calibration in a small box whose worldline
runs from $x_0$ to $y_0$
before performing the experiment. (See FIG. 4.)
However, this choice is vulnerable to quantum fluctuations.
Totally analogous to our discussion concerning FIG.\ 2
in Section 4, consider the spontaneous nucleation of
a virtual vortex-antivortex pair whose worldline is non-trivially linked to
the union of three object:
the Wilson loop, the tails and the string loop under calibration.
This will conjugate the measured flux relative to the calibrated value.
In the weak coupling expansion, such a configuration
has a single excited link on the path that connects $x_0$ to $\Sigma^*$.
This causes (in three spacetime
dimensions) the excitation of
four plaquettes and is suppressed by terms that are {\em independent\/} of the
size of $\Sigma^*$ and $C$ or the separation between them.
Thus,
higher order corrections render the flux uncertain
up to conjugation and
this operator is useless
as
an order parameter. This was the conclusion drawn by Alford {\em et al.}
\cite{qft} .

Such a conclusion is unwarranted
as it is based on an implicit choice of $\{ l_P \}$.
Before we present our resolution, let us note another related problem
that we have already raised in the last Section.
It is a subtle issue to maintain the {\em coherence} between various strings
when quantum fluctuations are taken into account.
With this implicit choice of long tails,
these quantum fluctuations do destroy such coherence.
This is because the calibrated flux of
one string may be conjugated while the elements associated
with others are unaffected.
This relative change in flux is highly physical and
does not go away even when we take the trace of our operator.

Consider the operator
\begin{equation}
 tr A^{\nu}_{a_1, a_2, \cdots, a_n} ( \Sigma^{*}_1, \Sigma^{*}_2, \cdots,
 \Sigma^{*}_n,x_0,\{l_P\};C) = { F_{a_1, a_2, \cdots, a_n}
( \Sigma^{*}_1, \Sigma^{*}_2, \cdots, \Sigma^{*}_n,x_0,\{l_P\}) W^{\nu}(C)
\over \langle F_{a_1, a_2, \cdots, a_n}
( \Sigma^{*}_1, \Sigma^{*}_2, \cdots, \Sigma^{*}_n,x_0,\{l_P\}) \rangle
\langle
W^{\nu}(C) \rangle} . \label{nstring}
\end{equation}
It was suggested in Ref. \cite{qft} that when a charged particle
winds around vortices of flux $a_n, a_{n-1}, \cdots, a_2, a_1$ in
succession, it will acquire an overall Aharonov-Bohm phase
\begin{equation}
 \lim ~ tr \langle A^{\nu}_{a_1, a_2, \cdots, a_n}
( \Sigma^{*}_1, \Sigma^{*}_2, \cdots,
 \Sigma^{*}_n,x_0,\{l_P\};C) \rangle =
\left({1 \over n_{\nu}}\right) \chi^{\nu}(a_1 a_2 \cdots a_n) .
\label{fcoherent}
\end{equation}
With our implicit choice of $\{l_P\}$ in FIG. 5, one finds,
contrary to the claim made in Ref. \cite{qft}, that (in three
spacetime dimensions) there is
a higher order correction term:
\begin{eqnarray}
 & & \lim ~ tr \langle A^{\nu}_{a_1, a_2, \cdots, a_n}
( \Sigma^{*}_1, \Sigma^{*}_2, \cdots,
 \Sigma^{*}_n,x_0,\{l_P\};C) \rangle \nonumber \\
&=&
\left({1 \over n_{\nu}}\right) \chi^{\nu}(a_1 a_2 \cdots a_n)
  +
\sum_g  O( \exp(8 \beta ({\text Re} \chi^{(R)}(g)- n_R)) )
\chi^{\nu}(ga_1g^{-1} a_2 \cdots a_n)
 ,
\label{incoherent}
\end{eqnarray}
where $n_{\nu}$ ($n_R$) is the dimension of the representation $\nu$ ($R$).
The second term on the r.h.s. (the higher order correction term)
shows that coherence of the strings
has been spoiled. (Cf. Eq.\ (\ref{pha}).)
Taken at face value, our results seem to suggest
that, because of quantum fluctuations,
construction of order parameters
for non-Abelian gauge theories is a hopeless enterprise.
Of course, this conclusion is only an
artifice of the particular choice of
$\{l_P\}$.
Let us look at the problem more closely.
Since conjugation of some (but not all) of the plaquettes
of a string requires the dynamical propagation of
strings carrying the commutator of the various inserted fluxes,
it is reasonable
to believe that configurations of this type are energetically costly.
For this reason, only conjugation of the calibrated flux of a whole
string loop deserves attention.
Here comes the question.
With an ingenious choice of $\{l_P\}$, can
one prevent vacuum fluctuations from conjugating the inserted flux of
a whole string loop
at a low energetic cost? The answer is
no. Since all the tails, $\{l_P\}$,
originate from the base point, quantum fluctuations can always conjugate
the flux of a whole string loop just by flipping all the links
from which the tails leave the base point.
This corresponds to the picture in which a virtual string loop
wraps around all the calibrating and measuring particles.
Fortunately, this only leads to a redefinition of the basis that
we
use for both calibration and measurement and does not affect
the result of our experiment.

The relevant
question really is: Are there choices of
$\{l_P\}$ by which one can prevent energetically
inexpensive vacuum fluctuations from conjugating
the inserted flux of a whole string loop {\em without} affecting the Wilson
loop? The answer is yes.
As emphasized in Section 3, any experimental determination
of the non-Abelian
Aharonov-Bohm factor essentially
proceeds in two stages: calibration and
measurement. Unless the two are done in a coordinated manner,
it is entirely understandable that one may be fooled by
quantum fluctuations. The idea is that
quantum fluctuations in FIG. 4 conjugate
the calibrated flux without
affecting the measuring apparatus, thus preventing the recovery of
the calibrated flux in the measurement. One can also consider
an analogous process in which quantum fluctuations affect the
measuring apparatus but not the calibrating apparatus. This would
correspond to the configuration depicted in FIG. 6 where the worldline of
the vortex-antivortex pair winds around the Wilson loop.
Note that the configuration in FIG. 6
is, in fact, a smooth deformation of that of FIG. 4.
In both figures, the vortex-antivortex
worldline has a non-trivial
linking number with the union of the Wilson loop , $\{l_P\}$. and the string
loop.

Recall that in the original choice of $\{l_P\}$ by Alford {\em et al.},
all the tails run along a chain of links from
$x_0$ to $y_0$ which is not on the Wilson loop.
Physically, this essentially means
that the particles used for calibration and the subsequent measurement
are kept in separate boxes. (We can regard a portion of
the Wilson loop as the box for storing particles
for subsequent measurement.) The problem is:
a virtual string loop may nucleate, wrap
around one of the boxes and re-annihilate. If this happens, the state
of charged particles contained in the box wrapped around by the string
loop
will change relative to those in the other box. Since we are using particles
in a particular box for calibration and those in the other box for
subsequent measurement. Clearly, we get different answer for the two
experiments. This is why this choice of  $\{l_P\}$ does not work.

\subsection{The Resolution}
Having observed this point, the resolution is simple.
We shall first present our resolution from a mathematical
point of view and then back it up with physical intuition.
First, note that
if the Wilson loop and the path of calibration were the same, i.e. $
C=l_P P l_P^{-1}$ for some $P$,
the Wilson loop would trivially recover the calibrated element.
A moment of thought reveals
that the key problems are: on the one hand, all the tails must merge
to the base point; on the other hand, since we take the limit that
the closest approach of $C$ to $\Sigma^*$ approaches infinity
in the definition of the order parameter (i.e., long range experiments),
the tails
${l_P}$'s  inevitably contain links that are not on the Wilson
loop.
Quantum mechanical fluctuations of those links affect the
calibration apparatus but not the measurment apparatus.
The original construction is particularly vulnerable because
there is a single link that belongs to all tails but is not on
the Wilson loop.
Branching may help reduce its vulnerability. Moreover, it may
be a good idea for at least
some of the tails to be initially on our Wilson loop, even though
they must eventually branch out from it.
Consider
the configuration shown in FIG. 7 where
the tails are chosen in such a way that many of them beginning from
the base point
are on our Wilson loop initially and branch out one by one
from it.
In  what follows, we argue that this construction
overcomes all difficulties caused by quantum fluctuations.
In order for quantum fluctuations to affect the calibration but not
the measurement, links on the Wilson loop must not be excited. Since
by construction the tails that branch out from the Wilson loop never
intersect one another after their branching out, to achieve
overall conjugation of the flux of string loops, we must then flip
a link in each tail after it has branched out.
Since the number of tails that branch out goes to infinity as
$\Sigma^*$ and
$C$ get large, we must excite a large number of links.
(See FIG. 7.) Such configurations have large actions and their
contribution to the partition function is suppressed by factors
proportional to the system size (e.g., the length of the Wilson loop).
Our conclusion is that
there is no energetically inexpensive
way of conjugating
the flux of a whole string loop without affecting
the Wilson loop.

The above paragraph requires a severe qualification.
It is important for our choice of the Wilson loop
not to come close to retracing itself.
Otherwise, it is possible for quantum fluctuations to affect the measuring
but not the calibrating apparatus. Suppose the Wilson loop is chosen to be of
the shape of a tennis racket. i.e. it runs
along a long chain $L$ of links from the base point $x_0$ to
a point $y_0$ and goes around various vortices of interest before
coming back to $y_0$ where it retraces $L$ back to $x_0$.
The tails are chosen such that many of them initially follow
the chain $L$ of links on the Wilson loop, but finally branch out
from $L$ one by one.
Now the Wilson loop is vulnerable to quantum fluctuations: A worldline
of virtual vortex can be linked to the Wilson loop near to $y_0$.
With the choice of the Wilson loop in FIG. 8, the worldline of the
virtual vortex, and therefore the energetic cost
one pays in conjugating the measured flux relative to its calibrated
value, is small.

There are strong physical motivations for our new choice of $\{l_P\}$.
That the tails initially follow the Wilson loop
corresponds to the physical picture that we keep
the charged particles that we use for {\em both} calibration and subsequent
measurement in the same box. The portion of the Wilson
loop that the tails initally lie on corresponds to the worldline of
this box. The fact that we are calibrating
every plaquette through which the string passes (stage one) means that we
calibrate the string loop continuously over time from the base point.
(We need to specify the whole worldsheet.)
Since the particles for both calibration and
subsequent measurement are stored in the same box,
lots of tails branch out from the Wilson loop.

Moreover, it is crucial not to send out the two beams of particles
for subsequent measurement (stage two) too closely spaced in time.
Otherwise, (with the
tennis racket choice in FIG. 8) quantum
fluctuations will spoil our result:
a virtual string loop
may wrap around both beams. This will conjugate the
measured flux.
To avoid this phenomenon, a wave packet
of some
test particles (one of the two beams)
should be kept in the box that store the particles (for
both calibration and measurement) and interfered
with another wave packet (the other beam) that traverses the string.
This is the physics underlying our new choice of $\{ l_P \}$.

Recall in the discussion in the last Section that it
is always possible for a virtual string loop to wrap around just one
of two beams that we use for subsequent measurement.
(This is true even in the case of an Abelian gauge theory.)
However, this incoherent effect will go away on average
if we repeat independent identical experiments many times to
extract the expectation value. (See Eq.\ (\ref{abelian})
in Section 3 and Eq.\ (\ref{correctop}) in
Section 7.)

In conclusion,
our construction coordinates the calibration and measurement processes,
making it impossible for our results to be sabotaged by just
a few small virtual string loops.
As noted earlier, it is always possible for a virtual string loop
to wrap around the box storing all the particles. This leads to no
real change as calibration and measurement are affected in the same
way.
The calculation in the next subsection will verify our assertion that
a
careful coordination between calibration and measurement solves the
problem of quantum fluctuations that has plagued all previous attempts
to construct order parameters for non-Abelian gauge theories.

\subsection{Calculation}
Having specified our choice of tails, we shall now prove that
at weak coupling,
Eq.\ (\ref{order})
is unspoiled by higher order corrections.
Let us first consider the vacuum expectation value of the
untraced Wilson loop operator
$\langle U^{(\nu)}(C, x_0)\rangle $.
Totally analogous to the $Z_2$ gauge-Higgs theory,
at weak couplings discussed in the Appendix, we
can completely ignore the matter
action without changing our conclusion.
Besides, the link excitations are so heavily suppressed that
our result will be dominated by configurations with a low density of
excited links. Therefore, configurations where
any two excited links share a plaquette can be
safely ignored,
thus enabling us to factorize the sum over the links, $U_{ij}$, into two
parts: one over the links that are on $C$ and one over that are not.
The contributions
from the second part are cancelled by a
corresponding sum in the denominator.
Now the gauge action, being gauge invariant, can be expressed in
terms of character functions and we obtain
\begin{equation}
\langle U^{(\nu)}(C, x_0) \rangle = {1 \over Z} \sum_{V_1} \sum_{V_2}
 \sum_{V_n} D^{(\nu)} (V_1 V_2 \cdots V_n) \times \prod_{i=1}^n
\sum_{\mu^{*}} C^{\mu^{*}}(\beta) \chi^{\mu^{*}} (V_i),
\label{exwilson}
\end{equation}
where $Z$ is same as the numerator but without the factor $D^{(\nu)}$.
As discussed before, when $F_a(\Sigma^{*}, x_0, \{l_P\})$ is inserted,
the configuration at weak couplings that has none of
its plaquette frustrated is an ``a-forest'' in which, roughly speaking,
all the links
that intersect the minimal surface with boundary $\Sigma^*$
are of flux $a$. (FIG. 3.) Suppose the Wilson loop
links once with the string loop. Exactly one of its links, $l$
(say with a flux $V_l$), is an
$a$-forest link. So, its parallel neighbors are all of flux $a$ and
any plaquette containing $l$ and one of its parallel neighbor
will be of flux $a^{-1}V_l$. We can take this into account simply
by defining a new variable $V'_l =a^{-1}V_l$. Thus,
in Eq.\ (\ref{exwilson}) we replace
$ D^{(\nu)} (V_1 V_2 \cdots V_n)$ by $ D^{(\nu)} (V_1 V_2 \cdots V_{l-1}a
V'_l \cdots V_n)$ and $V_l$ by $V'_l$ elsewhere.
It follows from the orthogonality relations between matrix elements
of irreducible representations,
\begin{equation}
 \int dU D^{(\nu)}_{ij} (U)  D^{(\mu)*}_{kl} (U) = {1 \over n_{\nu}}
\delta^{\mu \nu} \delta_{ik} \delta_{jl}, \label{ortho}
\end{equation}
that Eq.\ (\ref{order}) survives higher order corrections.

\subsection{More on Virtual Vortex Loops}
It is important to note that the a-forest, which is bounded by an inserted
string worldsheet, is unphysical
and can be moved around.
We leave it as an exercise for the reader
to show that the effect of the virtual vortex loop shown in
FIG. 9
can be safely ignored. Note that the virtual
vortex loop has trivial linkage with
the union of the Wilson loop, the tails and the inserted string.

Incidentally, a careful choice of $\{l_P\}$ and the Wilson loop
$C$ taken together will also maintain the quantum coherence between
various string loops. A bad choice of $C$ by which quantum coherence
is spoiled is shown in FIG. 10. A virtual vortex worldline
conjugates the measured flux of one vortex relative
to another. As a comparison, in FIG. 11, we show a choice of $C$ which
would preserve
quantum coherence. Let us consider paths which begin and end at points
on the Wilson loop. Some of those paths (e.g. $P_{pq}$) are contractible.
Some (e.g. $P_{yz}$) are not. We can classify paths
into classes in which they are smoothly deformable into one another.
Let us consider the
topologically non-trivial classes of paths. If the minimal lengths of all
such paths diverge as $\Sigma^*_1$,
$\Sigma^*_2$ and $C$ get large, then coherence of the inserted flux
between the two vortices can be maintained.
This is because quantum fluctuations that would conjugate the
flux of one string relative to another are energetically costly and
can, therefore, be safely ignored. Consequently,
for a charged particle traversing $n$ coherent
strings in succession, we obtain
\begin{eqnarray}
\langle A^{\nu}_{a_1, a_2, \cdots, a_n}
( \Sigma^{*}_1, \Sigma^{*}_2, \cdots,
 \Sigma^{*}_n,x_0,\{l_P\};C) \rangle
 &=&{ \langle  F_{a_1, a_2, \cdots, a_n}
( \Sigma^{*}_1, \Sigma^{*}_2, \cdots, \Sigma^{*}_n,x_0,\{l_P\}) U^{\nu}(C)
\rangle
\over \langle F_{a_1, a_2, \cdots, a_n}
( \Sigma^{*}_1, \Sigma^{*}_2, \cdots, \Sigma^{*}_n,x_0,\{l_P\}) \rangle
\langle
tr U^{\nu}(C) \rangle} \nonumber \\
 &=&
\left({1 \over n_{\nu}}\right) \chi^{\nu}(a_1 a_2 \cdots a_n) .
\label{trace}
\end{eqnarray}
\section{Gauge fixing}
We have considered
a discrete non-Abelian gauge group $G$ which arises as a result of
the spontaneous symmetry breakdown of a simply-connected
group $K$.
We have successfully
proved that in a free $G$ charge phase, with a careful choice of $\{l_P\}$,
Eq.\ (\ref{order}) remains valid
even when quantum fluctuations are fully taken into account.
One might naively expect that
the operator $A^{\nu}_{a}
( \Sigma^{*} ,x_0,\{l_P\};C)$ in
Eq.\ (\ref{order})
is the order parameter for non-Abelian gauge theories.

However, there
are still complications in the study of the symmetry breakdown:
$G \to H$. For one thing,
with the breaking of $G$ into $H$, an irreducible representation
$\nu$ of $G$ is typically reducible in $H$.
In general, only one of the irreducible
representations of $H$ that is contained in $\nu$ will dominate
the asymptotic behavior of $W(C)$ when $C$ is large.
Since
particles
in different irreducible representations of $H$ in the decomposition
can be resolved from one another, it makes no sense to normalize them
by the same factor $tr U^{\nu} (C, x_0)$ in the denominator
of the first line of Eq. (\ref{order}).
The correct thing to do is to consider each irreducible representation of
$H$ separately and normalize
the matrix elements (for each irreducible representation) by dividing the trace
over the particular representation that one is working with.

There is a more serious problem.
Unless $H$ is in the center of $G$, it does not make any
sense to talk about irreducible representations of $H$ without
some gauge fixing scheme.
In particular, when $G$ is spontaneously broken into a non-normal subgroup $H$,
it is easy to see that our original set of order parameters does not
work. A gauge transformation by $g \in G$ at $x_0$ in
Eq.\ (\ref{order}) shows that
\begin{equation}
A^{\nu}_{ghg^{-1}}
( \Sigma^{*}, x_0,\{l_P\};C) \rangle =U^{\nu}(g ) A^{\nu}_{h}
( \Sigma^{*}, x_0,\{l_P\};C) U^{\nu}(g^{-1}).
\label{dependent}
\end{equation}
The order parameters $A^{\nu}$ for $h$ and $ghg^{-1}$ are not independent
and there is no way for us to distinguish the behavior of a string
$h \in H$ with one $ghg^{-1} \notin H$.
The way out is to perform gauge fixing. (To be more
precise, the embedding the unbroken group $H$ in $G$ changes as $\phi$ varies.
One should take this into account and consider $H$ as a function of
$\phi$, i.e., $H(\phi)$.) Suppose
$G$ acts transitively
on the Higgs $\phi$.
(We believe that the requirement that $G$ acts transitively on $\phi$
is only a working assumption. For configurations of finite energy,
the Higgs field necessarily approaches its vacuum expectation value at spatial
infinity and our gauge fixing procedure is well defined.)
Without loss of generality, one can consider
the gauge fixed insertion operator, $F_a(\Sigma^{*}, x_0, \{l_P\},\phi_0)$,
where the Higgs field $\phi= \phi_0$. When this operator is
inserted in a Green's function
with gauge invariant operators, its gauge invariant part will be
projected out. Thus, it will have the same effect as the gauge invariant
operator, $ {1 \over |H|} \sum_{h \in H(\phi)}
F_{hah^{-1}}(\Sigma^{*}, x_0, \{l_P\},\phi)$.

Even in the case that $H$ is normal in $G$ but not in the center of $G$,
it is still {\em necessary\/} to gauge fix
$\phi = \phi_0$ for the untraced Wilson
loop operator $U^{\nu} (C)$.
The reason is that,
without gauge fixing, $U^{(\nu)} (C, x_0) \to D^{(\nu)}(g)
U^{(\nu)} (C, x_0)  D^{(\nu)}(g^{-1}) $ under
a global gauge transformation by $g$. The expectation
value $\langle U^{(\nu)} (C, x_0) \rangle $ is invariant under $G$.
By the Schur's lemma, $\langle U^{(\nu)} (C, x_0) \rangle = \lambda I$
for some $\lambda$. The interpretation is clear:
an irreducible representation of $G$ is typically reducible in
$H$. In the absence of
gauge fixing, it would not be possible for us to
resolve the
various irreducible
representations of $H$ in the decomposition.

Therefore, we should always gauge fix $\phi=\phi_0$ and consider
$U^{(\nu)} (C, x_0, \phi_0)$. From now on, we will only
be interested in gauge transformation by
the unbroken group $H(\phi_0)$. Let us decompose the representation
$(\nu)$
into irreducible representations $(\mu_1), (\mu_2) , \cdots ,(\mu_n)$
of $H$ and the representation space $V= V_1 \oplus V_2 \cdots \oplus V_n$.
The leading non-trivial contribution to $\langle U^{(\nu)} (C, x_0, \phi_0)
\rangle$ must arise when one of the links on $C$ takes a value $h$
in the unbroken group $H$. (If it takes a value $g \notin H$, some matter
fields are excited. Thus, its contribution to
$\langle U^{(\nu)} (C, x_0, \phi_0) \rangle$ must be suppressed.)
By gauge fixing $\phi= \phi_0$, we can disentangle
various irreducible representations of $H$
from one another and
see clearly that the symmetry group has been
reduced to $H$.
In other words, in the free $H$ charged phase and with gauge fixing, we obtain
again Eq.\ (\ref{order}) for a representation of $H$ (rather than $G$).
As discussed in Section 4, a string with flux $a \notin H$ is now bounded
to a domain wall which decays to a composite string with flux $ab^{-1} \in H$.
Thus, the Aharonov-Bohm phases will give elements of $H$ only
and this shows clearly that the unbroken group is $H$.
{}From that point on, there is no obstacle to repeating our original
analysis
to study any symmetry breaking of $H$ at an even lower energy scale.

\section{Discussions}
\subsection{Gauge-invariant Formulation}
Note that the operators
$ F_a (\Sigma^{*}, x_0, \{l_P\},\phi_0 )$
and $U^{\mu}(C, x_0, \phi_0)$ we use in the definition of our order
parameters are not gauge invariant under $H$.
However, it is possible for
us to redefine the order parameters in terms of
quantities which are invariant under $H$.
This is done by tracing over
an irreducible representation $(\mu)$ of $H$ in the Wilson loop operator and
for the insertion of strings, we use the
operator
$ {1 \over |H|} \sum_{h \in H(\phi_0)}
F_{hah^{-1}}(\Sigma^{*}, x_0, \{l_P\},\phi_0)$.
In conclusion, we propose that
\begin{equation}
\langle A^{\mu}_{a}
( \Sigma^{*}, x_0,\phi_0, \{l_P\};C) \rangle
 ={ \left\langle {1 \over |H|} \sum_{h \in H(\phi_0)}
F_{hah^{-1}}(\Sigma^{*}, x_0, \{l_P\},\phi_0 ) tr U^{\mu}(C, x_0, \phi_0)
\right\rangle
\over \left\langle {1 \over |H|} \sum_{h \in H(\phi_0)}
F_{hah^{-1}}(\Sigma^{*}, x_0, \{l_P\},\phi_0 ) \right\rangle
\left\langle
tr U^{\mu}(C,x_0, \phi_0) \right\rangle}
\label{correctop}
\end{equation}
is an
order parameter for non-Abelian gauge theories.
At first sight, our order parameters
may appear to be redundant because, their definition
seems to require a knowledge of the identity of the Higgs field and
the action
of the gauge group on it. Doesn't it mean that we
already know what the unbroken group is?
Our view point is that
one should think of $\phi$ only as a {\em candidate} for the
Higgs field rather than the Higgs field itself. We are {\em testing}
with our order parameters
if a Higgs phenomenon has occurred.
The actual realization of the gauge
symmetry depends on the values of the coupling constants in our theory.
\subsection{Example: $S_3 \to Z_2$}
Since the above discussion is rather abstract, one desires an explicit
model in which the bahavior of the order parameter $A^{\nu}_a (\Sigma, C)$
can be studied analytically. However, there are some technical
difficulties. A matter action on a lattice
that we find tractable is
of the form
\begin{equation}
S_{\rm Higgs} = - \sum_{\mu} \gamma_{\mu} \sum_l \left( \chi^{\mu} ((
\phi^{-1} U \phi)_l) + c.c. \right),
\label{tract}
\end{equation}
where the sum runs over all irreducible representations of $G$. Perturbative
methods can be used to analyze this model when some of the $\gamma_{\mu}$'s
are large ( $ \gg 1$) while all others are small ($ \ll 1$ ).
In the weak coupling limit $\gamma_{\mu} \to \infty$, $\phi^{-1} U \phi$
becomes restricted to the kernel of the representation $( \mu )$ at each
link and $U_P$ takes value in the kernel at each plaquette. In other
words, $G$ is spontaneously broken into the subgroup $H = Ker( D^{\mu})$,
which is normal in $G$.
Indeed, the breaking of $G$ into any {\em normal} subgroup can be
obtained by choosing some $\gamma_{\mu}$'s to be large.
Of course, it is still conceivable that at immediate couplings, a
more general symmetry breaking pattern can occur and a non-normal
subgroup may happen to be the unbroken group. However, at immediate
couplings, our perturbative methods clearly break down and there
is no simple way of analyzing the result other than numerical methods.

It would be very helpful if one can come up with a
more general tractable lattice action
in which a non-Abelian group is spontaneously
broken into a {\em non-normal} subgroup.
Let the matter field $\phi$ take
values in the left coset of $H$ (denoted by $G/H$)
and consider the action
of the form
\begin{equation}
S= -\beta  \sum_P ( \chi^{R} (U_P) + c.c.) - \gamma \sum_l F
\left( \phi_i , U_{ij}
\phi_j \right) ,
\label{discretenn}
\end{equation}
where
$F$ is a mapping from $G/H \times G/H$ to real numbers.
$F$ has to respect gauge invariance. i.e.
$F( gg_1 H, gg_2 H)= F( g_1 H,g_2 H).$
Besides, $F(H, H) \geq F(g_1H, g_2 H)$ for all $g_1, g_2 \in G$.
Let us apply our formalism to the case of the
symmetry breakdown of $S_3$ to $Z_2$.
Without loss of generality,  let $H =\{e, (12) \}$. Suppose
$F(H,H) =1$ and $F \left(H, (23)H \right) = F \left(H, (13) H \right) =0$.
We shall assume without proof that there exist convergent
weak and strong
coupling expansion schemes for the matter action of Eq.\ (\ref{discretenn}).

(1) $\beta \gg 1, \gamma \ll 1 $

Just like the $Z_2$ gauge-Higgs system discussed in
the Appendix, the matter action can
be safely ignored.
$ \langle F_a(\Sigma^*) \rangle $ is dominated by
an $a$-forest. Our order parameters show that the theory is
in a free $S_3$ charge phase.

(2) $\beta , \gamma \gg 1$

Now the leading non-trivial contribution to $ \langle W(C) \rangle$
arises when one of the links on $C$ has $U_l = (12)$. (A configuration
with one of the links on $C$ taking a value $U_l \not= e$ or $(12)$
excites the matter field and its contribution is thus severely
suppressed.)
It is easy to check that the two-dimensional irreducible representation
of $S_3$ is decomposed into a trivial and a non-trivial representations
of $Z_2$. With gauge fixing, the two representations
can easily be resolved from each other.
While $F_{(12)}$ inserts a stable string loop, other $F_a$ insert
string loops which are bounded by domain walls.
Thus, the
theory is in a free $Z_2$ charge phase.

{}From the non-analytic behavior of the order parameters across the
phase boundary, we see that there is a symmetry breaking from
$S_3 $ to $Z_2$.
\subsection{Vacuum Overlap Order Parameters}
Other order parameters for gauge theories
have also been previously proposed. (For a review,
see Ref.\ \cite{Freden}.)
One promising probe of the phase structure of a gauge theory
is the
vacuum overlap order parameters (VOOP) proposed by
Fredenhagen and Marcu \cite{FreMar}.
Suppose a matter field $\Phi^{(\mu)}$ which tranforms irreducibly
under a gauge group $G$. Choose a path $P_{x,y}$ which connects two
widely separated points $x$ and $y$ and consider the gauge-invariant
operator
\begin{equation}
K^{(\mu)} (x, y , P_{x,y}) = \Phi_x^{(\mu) \dagger} D^{(\mu)}
\left( \left( \prod_{l \in P_{x,y}} U_l \right) \right) \Phi_y^{(\mu)}.
\label{VOOP}
\end{equation}
If the gauge group $H$ is unbroken, the field $\Phi^{\dagger}$ should
create a stable particle which will propagate between $x$ and $y$.
Therefore, we have
\begin{equation}
\langle K^{(\mu)} (x, y , P_{x,y}) \rangle
\sim \exp \left( -M^{(\mu)}_{\rm ren} L (P) \right) \exp
\left( - M^{(\mu)}_{\rm dyn}|x- y| \right) ,
\label{vVOOP}
\end{equation}
where $M^{(\mu)}_{\rm ren}$ is the renormalised mass of the classical source
of charges propagating along $P$,
$M^{(\mu)}_{\rm dyn}$ is the dynamical mass
of the stable particle created by $\Phi^{(\mu) \dagger}$,
$L(P)$ is
the length of $P$ and $|x -y|$ is the distance from $x$ to $y$.
If the representation $(\mu) $ is confined or screened, there will
be no dynamical propagation of stable charged particles. Thus, we expect
that $\langle K^{(\mu)} (x, y , P_{x,y}) \rangle$ is independent of
$|x -y|$ for large separation. i.e. $M^{(\mu)}_{\rm dyn} =0$.
Fredenhagen and Marcu proposed that $M^{(\mu)}_{\rm dyn} > 0$ is
the criterion for $ \Phi^{\dagger}$ to create a free charge.
Their construction is highly similar to our order parameters.
When the gauge group is broken into a subgroup $H$,
the complications in disentangling various irreducible representations
of $H$ in the decomposition of an irreducible representation of $G$
discussed in the last section also arise here.
We expect the resolution is again gauge fixing. It would be
interesting to work it out.
Finally, we remark that application of the VOOP to the
study of partial symmetry breakdown has been considered in the explicit
example of the Georgi-Glashow model \cite{Filk}.
\subsection{Phase Transition Without Symmetry breaking?}
It was suggested in Ref. \cite{qft} that there is a possibility
of having a phase transition without a change in the symmetry group.
Recall that generally just one irreducible representation of $H$ will
dominate the asymptotic behavior of $U^{\nu}$.
Alford {\em et al.} proposed
that in some parameter space of the theory
a ``crossover'' may occur, where this
representation changes. We are not sure whether such an
interesting phenomenon is possible.
\subsection{Conclusions}
Our order parameters are useful for studying the symmetry
breakdown of non-Abelian gauge theories.
While we have concentrated our discussion on discrete group theories,
we emphasize that the idea of the Aharonov-Bohm
order parameter is rather general. The assumptions of
the existence of topologically stable flux tubes and the symmetry
group being discrete may be relaxed.
The subtleties of quantum fluctuations and gauge fixing are
intrinsic properties of non-Abelian gauge theories.

We emphasize that, after gauge fixing, vortices
of conjugate but different fluxes should be regarded
as non-identical. The coherence of the fluxes between
various strings is characteristic of non-Abelian gauge theories.
Quantum fluctuations tend to destroy these two important
features. In our construction of the order parameter,
we show how these problems can be overcome by a
careful coordination between the calibrating and
measuring processes.

When a gauge group $G$ is broken
into $H$, an irreducible representation of $G$ is typically reducible in $H$.
Particles in different
irreducible representations of the decomposition
can be resolved from one another.
In order to isolate the behavior of a particular irreducible representation
of $H$, it is crucial to gauge fix the Higgs field $\phi= \phi_0$.
This simple but crucial point has also been largely overlooked in previous
works.
We also sketch briefly the
application of the Aharonov-Bohm
order parameters to study the symmetry breaking of
$S_3$ to $Z_2$ and discuss the vacuum overlap order parameters
suggested in
the literature.

We would also like to remark that a Chern-Simons term can be added to
the action of a gauge theory in $2+1$ dimensions.
Our construction remains to be generalized to this
case. It is also of interest to note that linked Wilson
loops
are useful as order parameters for a Maxwell-Chern-Simons-Higgs
system. In the unbroken Chern-Simons phase,
matter charges are attached with fluxes, thus experiencing
Aharonov-Bohm interactions with one another. There is,
however, no such Aharonov-Bohm interactions in the Chern-Simons Higgs phase
because the Higgs mechanism removes the fluxes that are attached to the matter
charges in the unbroken phase \cite{prop}.
Finally, from a mathematical point of view, it is conceivable that
these types of
non-local objects
involved in the construction of order parameters
may give rise to interesting topological
invariants \cite{Witten}.
\section*{acknowledgments}
We are indebted to J. Preskill for bringing the problem
of quantum fluctuations in the context of order parameters for
non-Abelian gauge theories to our attention
and for stimulating discussions at
various stages of this work. Numerous helpful conversations with S.
Adler, M. Bucher,
H. F. Chau,
K.-M. Lee, J. March-Russell, P. McGraw, F. Wilczek,
M. de Wild Propitius and P. Yi are also gratefully acknowledged.
This work was supported in part by DOE DE-FG02-90ER40542.
\appendix
\section*{Example: $Z_2$ gauge-Higgs system}
Here we review the application of the
order parameter to a simple
model in the work of Preskill and Krauss \cite{PreKra}:
$Z_2$ lattice gauge theory coupled to a $Z_2$ spin system.
The degrees of freedom of the model are gauge variables
\begin{equation}
U_l \in Z_2 \equiv \{1, -1 \} ,
\label{gauge}
\end{equation}
residing on links (labeled by $l$) of a cubic four-dimensional
spacetime lattice and spin variables
\begin{equation}
\phi_i \in Z_2 \equiv \{1, -1 \} ,
\label{spin}
\end{equation}
residing on sites (labeled by $i$).
The partition function of the theory is
\begin{equation}
Z = \sum_{\{U\} \{\phi\}} e^{-S} ,
\label{partition}
\end{equation}
where the Euclidean action is
\begin{equation}
S= S_{\rm gauge}+ S_{\rm spin} ,
\label{aaction}
\end{equation}
where
\begin{equation}
S_{\rm gauge}= - \beta \sum_P U_P,
\label{gaction}
\end{equation}
and
\begin{equation}
S_{\rm spin}= - \gamma \sum_l ( \phi U \phi)_l,
\label{saction}
\end{equation}
where $U_P = \prod_{l \in P} U_l $ associates with each elementary plaquette
$P$ the product of the four gauge variables $U_l$'s  sitting on its links,
and $( \phi U \phi)_{ij} = \phi_i U_{ij} \phi_j$, for each pair of
neighboring sites. The action is invariant under the $Z_2$ gauge
transformation defined by
\begin{equation}
\eta_i \in Z_2 \equiv \{1, -1\},
\label{gdegree}
\end{equation}
where the variables transform as
\begin{equation}
\phi_i \to  \eta_i \phi_i , U_{ij} \to \eta_i U_{ij} \eta_j .
\label{gauget}
\end{equation}

Note that the gauge variable $U_l$ is invariant under a non-trivial
{\em global}
gauge transformation $ \eta_i = -1$, but the spin variable $\phi_i$
is not. The spin variable is, therefore, a matter field with a non-trivial
$Z_2$ charge and we would like to determine if there is a $Z_2$ superselection
rule.

Now we must consider how the operator $F(\Sigma^*)$  is to be defined on a
lattice. Recall that inserting $F(\Sigma^*)$ into a Green's function is
supposed to be equivalent to introducing a classical cosmic string
source on the world sheet $\Sigma^*$. In a (3+1)-dimensional
lattice, we consider $\Sigma^*$ to be a closed surface made up of
plaquettes of the {\em dual} lattice. There is a set $\Sigma$ of
plaquettes of the original lattice that are dual to the plaquettes of
$\Sigma^*$. (It is easier to visualize in three spacetime
dimensions. Then $\Sigma^*$ is a closed path made up of
links in the dual lattice. Each link of $\Sigma^*$
is dual to a plaquette of the original lattice.
See FIG. 3.) The operator $F(\Sigma^*)$ modifies the
gauge action of these plaquettes:
\begin{equation}
-\beta U_P \to \beta U_P, P \in \Sigma .
\label{Saction}
\end{equation}
This is equivalent to flipping the sign of $\beta$ in these plaquettes.

Let us consider a pure gauge theory first.
$\Sigma^*$ is the boundary of a set of cubes of the dual lattice.
The insertion of $F(\Sigma^*)$ is equivalent to performing
a singular gauge transformation
\begin{equation}
U_l  \to - U_l
\label{Faction}
\end{equation}
on all the links that are dual to those cubes.
(As shown in FIG. 3, in three spacetime dimensions, $\Sigma^*$
is the boundary of a set of plaquettes in the dual lattice. The insertion
of $F(\Sigma^*)$ is equivalent to performing singular gauge transformations
$U_l \to -U_l$ on all the links dual to these plaquettes. Those links
are marked by arrows.)
 So,
$F(\Sigma^*)$ is just a change of variable and
\begin{equation}
\langle F(\Sigma^*) \rangle =1 .
\label{Ftrivial}
\end{equation}

The Wilson loop on the lattice is defined as
\begin{equation}
W(C) = \prod_{l \in C} U_l,
\label{AWilson}
\end{equation}
where $C$ is a closed loop of links.
If the surface $\Sigma^*$ and the loop $C$ have a linking number 1,
$ F(\Sigma^*) $ flips the sign of one $U_l$ on $C$ and we find
\begin{equation}
\langle F(\Sigma^*) W(C) \rangle = - \langle W(C) \rangle .
\label{ztwo}
\end{equation}
This show that $Z_2$ charge is not screened in a pure gauge system
and a $Z_2$ cosmic string can be detected at long range by a $Z_2$
charge.

Let us now turn to the full theory: $Z_2$ gauge-Higgs system.
This model is tractable
because it can be analyzed by means of convergent perturbation expansions.
The phase structure of this theory shown
in FIG. 12 has previously been conjectured
and confirmed by Monte Carlo simulations. Preskill
and Krauss have shown that the order parameter $A(\Sigma, C)$ is an
appropriate order parameter.
To avoid overburdening the reader with technical details, we shall only
present explicit calculations in two regions. These calculations are
sufficient to prove their case.

(1) $\beta \gg 1$, $\gamma \ll 1$

In this region, $\exp (- \beta) $ and $\gamma$ are small. So, the gauge
variables are hard to excite but the spin variables are easy.
Therefore, configurations with only a small number of frustrated plaquettes
will be important and we can
expand the gauge action in powers of $\exp(-\beta)$. In effect,
we are expanding in terms of world sheets of virtual strings.
The matter action can be expanded, in powers of $\tanh \gamma$, as
\begin{equation}
e^{-S_{\rm spin}} = N(\gamma) \prod_l [ 1+ (\phi U \phi)_l \tanh \gamma ].
\label{mexpand}
\end{equation}
Consider the leading non-trivial contribution to $ \langle W(C) \rangle
$. It is zeroth order in $\tanh \gamma$ and
indifferent to spin frustrations. Therefore, we can
safely ignore the matter action without changing our conclusion.
Considering the gauge action alone, the leading non-trivial contribution
to  $ \langle W(C) \rangle$
arises when one of the
links on $C$ has $U_l = -1$. In four spacetime
dimensions, this will frustrate six
plaquettes that contain the link. (For ease of
visualisation, the corresponding picture in three
dimensions in which four
plaquettes are frustrated is drawn in FIG. 13. It corresponds simply to
the physical picture of having
a small virtual vortex worldline linked to the Wilson loop.) Thus, we find
\begin{equation}
\langle W(C) \rangle = { \exp (-L (e^{-2 \beta})^6 + \cdots )
\over \exp (+ L (e^{-2 \beta})^6 + \cdots )} =
\exp [-2 L (e^{- 2 \beta} )^6 + \cdots ],
\label{Wleading}
\end{equation}
where $L$ is the number of links on $C$. The exponentiation results
from summing over the $L^N / N!$ ways of flipping the sign of $N$ of the links
on $C$.
For $\langle F( \Sigma^*) \rangle$, the leading contribution
is obtained by flipping all the links dual to the volume enclosed by
$\Sigma^*$. Then $U_P = -1$ on the plaquettes dual to $\Sigma^*$ and
$U_P = 1$ elsewhere, so that no plaquette variables are frustrated.
Let us call this set of links with flux $-1$ a ``$-1$-forest''. (See FIG. 3.)
This $-1$-forest configuration, in which the gauge variables
$U_l$ are flipped in a volume bounded by $\Sigma^*$,
dominates $\langle F( \Sigma^*) \rangle$ because the gauge variables are
ordered and costly to excite, while the spin variables are disordered and
nearly indifferent to a flip in their nearest-neighbor couplings inside
$\Sigma^*$. We expand the spin partition function with the plaquette variables
frozen at these values to find
\begin{eqnarray}
\langle F( \Sigma^*) \rangle =
 { [ 1 - (\tanh \gamma)^4 ]^A \over [ 1 + (\tanh \gamma)^4 ]^A} \nonumber \\
 = \exp [ -2A (\tanh \gamma )^4 + \cdots ].
\label{Fleading}
\end{eqnarray}
Here summing over the spin variables around each plaquette on $\Sigma$ gives
a factor ${ [ 1 - (\tanh \gamma)^4 ] \over [ 1 + (\tanh \gamma)^4 ]}$ and
$A$ is the area of $\Sigma^*$.
With the contribution to $A(\Sigma^*, C)$ being dominated
by the $-1$-forest, a cosmic string has hair and we find
\begin{equation}
\lim A(\Sigma^*, C) = -1,
\label{Z2result}
\end{equation}
if $\Sigma$ and $C$ have an odd linking number.
Therefore, there is a $Z_2$ superselection rule. In other words,
$Z_2$ is the manifest low energy symmetry group.

(2) $\beta , \gamma \gg 1$
Once again the leading non-trivial contribution to
$\langle W(C) \rangle$ arises when one of the links on $C$ has
$U_l= -1$. The only difference is that flipping $U_l$ now frustrates
the spins on the link as well as the six plaquettes that contain the link.
Therefore,
\begin{equation}
\langle W(C) \rangle = \exp [ -2 L (e^{-2 \beta})^6 e^{-2 \gamma} + \cdots ],
\label{Wlead}
\end{equation}
where $L$ is the length of $C$.
The crucial difference from region (1) lies in the behavior of
$ \langle F( \Sigma^*) \rangle $. Since spin frustration is now costly,
the leading contribution to $ \langle F( \Sigma^*) \rangle $ no
longer arises from a $-1$-forest. A $-1$-forest will frustrate
spins in a {\em volume} bounded by $\Sigma^*$. The
preferable configuration is to frustrate all plaquettes dual to $\Sigma^*$.
This gives an area law decay
\begin{equation}
\langle F( \Sigma^*) \rangle =  ( e^{-2 \beta})^A + \cdots,
\label{Flead}
\end{equation}
where $A$ is the area of $\Sigma^*$.

Since the gauge variables $U_l$'s deep inside the volume bounded by $\Sigma^*$
are unaffected by the insertion of $ F( \Sigma^*)$, we see clearly
that
\begin{equation}
\lim ~\langle A(\Sigma^*, C)\rangle =1.
\end{equation}
The interpretation is simple. Spontaneous symmetry breaking of
$Z_2$ has occurred. Condensation of the matter field causes the string to
become the boundary of a domain wall, but the wall is unstable and decays
by nucleation of a loop of string. The inserted $Z_2$ string thus becomes
bounded to another $Z_2$ string and the composite object then gives
a trivial Aharonov-Bohm factor.

{}From the different behaviors in the two regimes, one concludes that
$\langle A(\Sigma^{*}, C) \rangle$ is an appropriate order parameter
for the $Z_2$ model. This result can be readily generalized to a lattice
gauge theory with an arbitrary Abelian gauge group.

\begin{figure}
\caption{(a) The test particles that we use for calibration and measurement
are stored in separated boxes. This arrangement is vulnerable to
quantum fluctuations. More concretely, a virtual string loop may nucleate,
wrap around one of the boxes and annihilate, thus changing the state
of the particles in one box but not the other. When we use
test particles in the two boxes to determine the flux of a string loop,
they give two different values (which are related by conjugation by
the group element associated with the virtual string loop).
(b) If the test particles for both calibration and subsequent measurement
are stored in the same box, the problem disappears as any virtual string
loop which affects the particles for calibration is going to affect
those for subsequent measurement in the same way.}
\label{fig1}
\end{figure}
\begin{figure}
\caption{Two consecutive  Aharonov-Bohm experiments are performed
to measure the flux of a vortex.
The dashed line (with arrow)
represents the worldline of a virtual vortex, which is
linked to the union of the two Wilson loops and the vortex worldline
of interest.
The second measurement of the flux gives a value which is a conjugate
of that of the first. i.e. quantum fluctuations render the flux of a
string uncertain up to conjugation.}
\label{fig2}
\end{figure}
\begin{figure}
\caption{The dashed line is $\Sigma^{*}$, comprised of links of the
dual lattice. The plaquettes shown belong to $\Sigma$ and are dual to
the links of $\Sigma^{*}$. The links marked by arrows are a-forest
links. i.e. In the leading weak coupling expansion, they are of flux
$a$.}
\label{fig3}
\end{figure}
\begin{figure}
\caption{Suppose all $l_P$ merge together at some point $y_0$ {\em not}
on the Wilson loop
before reaching the base point.
A worldline of virtual vortex conjugates the calibrated flux.}
\label{fig4}
\end{figure}
\begin{figure}
\caption{Suppose the $l_P$ from each vortex merge together before reaching
the basepoint. There
exists a short worldline
of virtual vortex (dashed line) which is topologically linked to the rest of
the figure. Owing to quantum fluctuations, the calibrated flux of vortex 2
relative to that of vortex 1
is rendered uncertain up to conjugation. This would destroy the coherence
of flux between the two vortices.}
\label{fig5}
\end{figure}
\begin{figure}
\caption{A deformation of the configuration shown in FIG. 6.
The worldline of the virtual vortex now winds around the Wilson loop,
thus affecting the measurement, but not the calibration apparatus.}
\label{fig6}
\end{figure}
\begin{figure}
\caption{Lots of long tails initially lie on the Wilson loop $C$.
They eventually
branch out from it one by one and never intersect one another afterwards.
Moreover, the Wilson loop never comes close to retracing itself.
To conjugate the calibrated flux without affecting the measurement,
each tail $l_{P_i}$
must be wrapped around by a virtual string loop
after its branching out from the Wilson loop.
Since the number of tails becomes large as $C$ and $\Sigma^*$ get large,
such configurations
are  energetically costly. We, therefore,
conclude that,
with a coordinated choice of the Wilson loop and $\{l_P\}$,
any energetically inexpensive excitation that affects the calibration process
necessarily affects the measurement process and vice versa.}
\label{fig7}
\end{figure}
\begin{figure}
\caption{The Wilson loop, $C$, is of the shape of a tennis racket with a long
chain running from $x_0$ to $y_0$. Starting from the base point, lots
of long tails initially lie on the long chain. They
ultimately branch out from it one by one and never intersert
one another thereafter. A virtual vortex worldline (the dashed line)
winds around the Wilson loop, thus affecting the measurement process
but not the calibration.}
\label{fig8}
\end{figure}
\begin{figure}
\caption{The effect of the virtual vortex worldline (the dashed line)
shown in the figure
can be safely ignored. The ``a-forest'' is unphysical and can be moved around
by gauge transformations. The virtual vortex worldline is unlinked to the rest
of the figure.}
\label{fig9}
\end{figure}
\begin{figure}
\caption{The Wilson
loop winds around vortex 2 first and vortex 1 second. Suppose
it is chosen such that it returns to the base point after winding around
vortex 2. A virtual vortex worldline of small size can conjugate
the measured flux of an inserted string relative to another.}
\label{fig10}
\end{figure}
\begin{figure}
\caption{The path $P_{pq} $ joining the points $p$ and $q$ on the Wilson
loop is contractible while there is topological obstruction to the shrinkage
of
the path $P_{yz}$ to a single point.}
\label{fig11}
\end{figure}
\begin{figure}
\caption{Phase diagram of the $Z_2$ gauge-Higgs system. \hfill}
\label{fig12}
\end{figure}
\begin{figure}
\caption{The leading non-trivial contribution to $\langle W(C) \rangle$
has a single link flipped ($U_l = -1$). Note that the Wilson loop and
the virtual vortex world line (denoted by the dashed line with arrow)
have a linking number one.}
\label{fig13}
\end{figure}
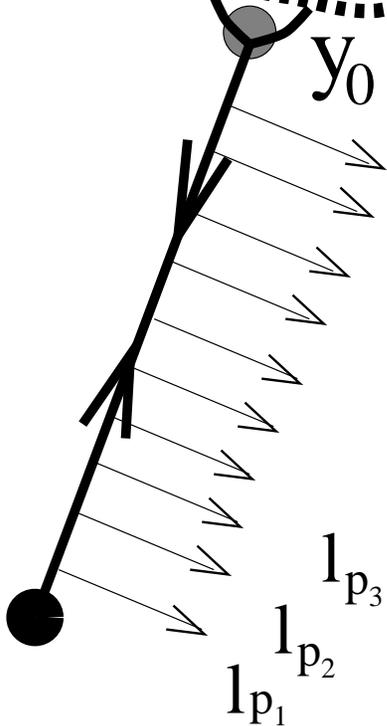
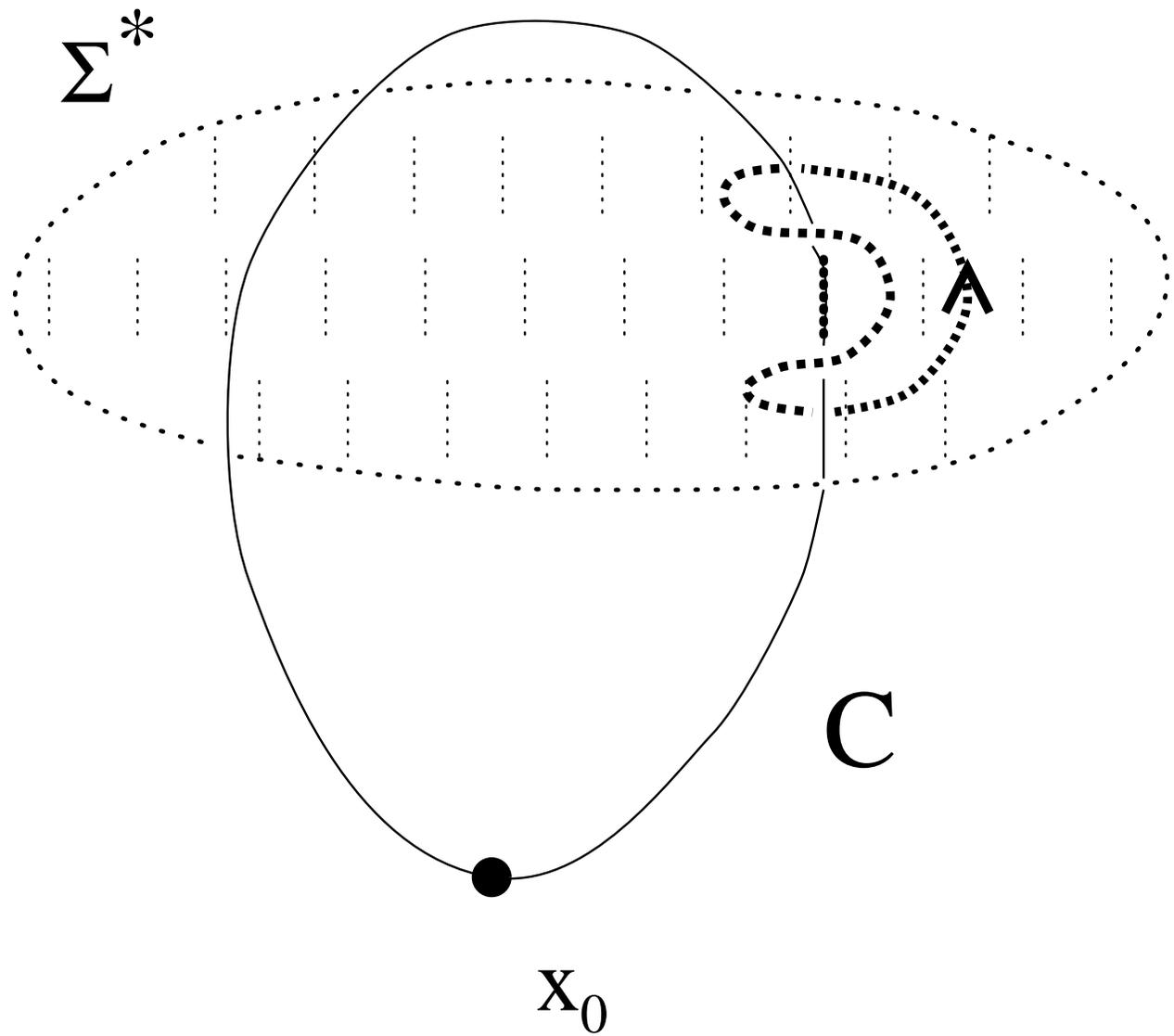
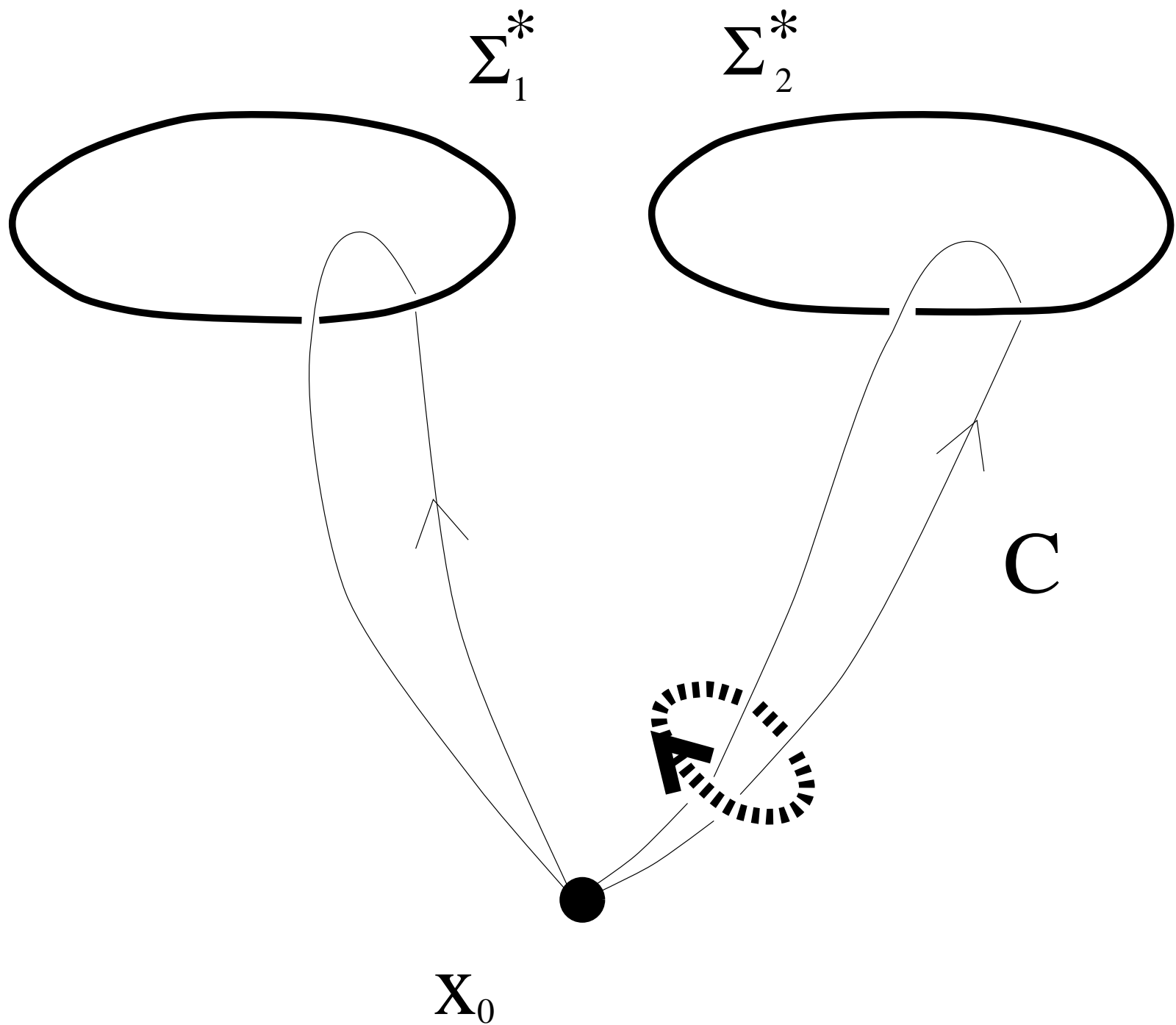
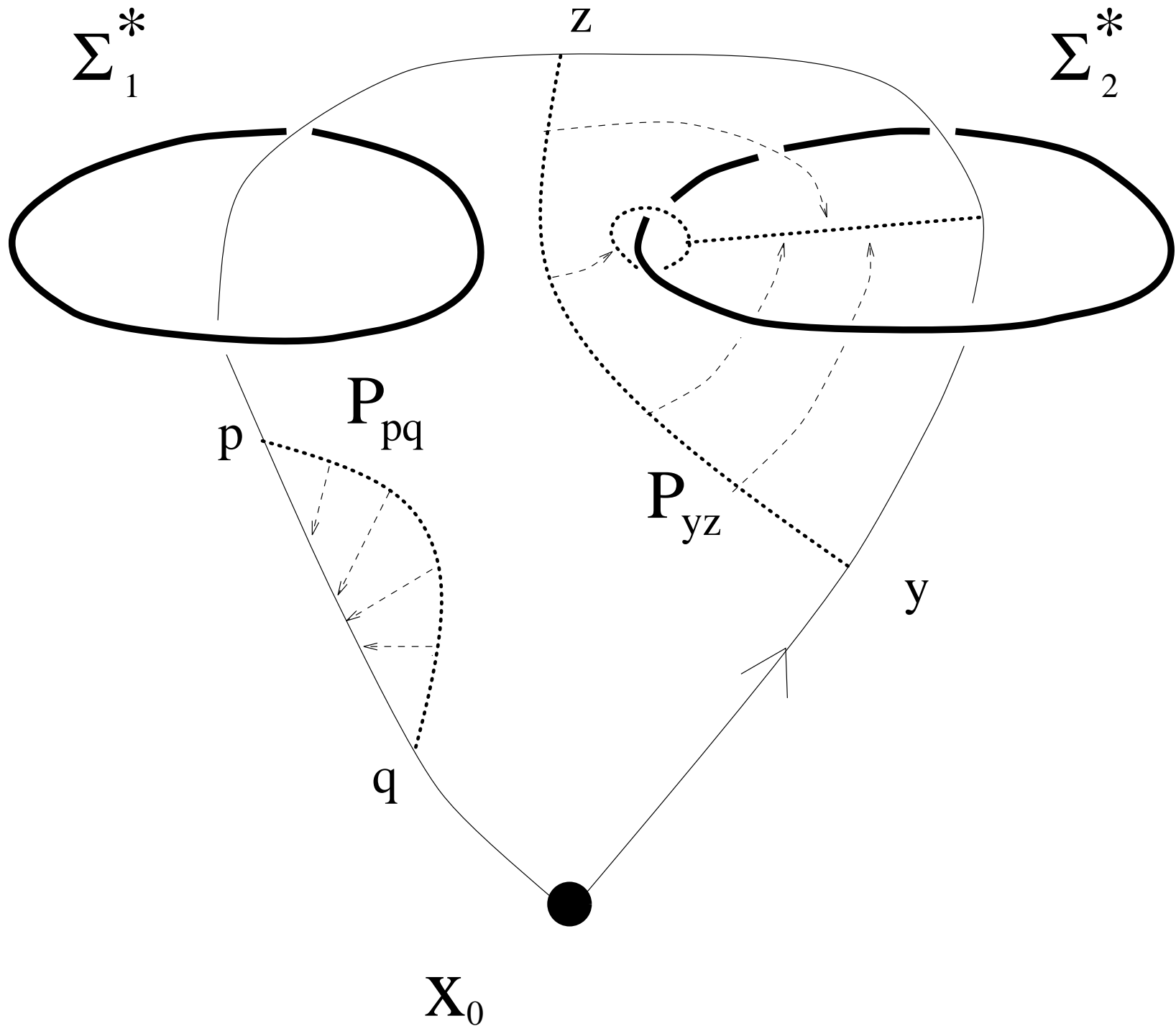
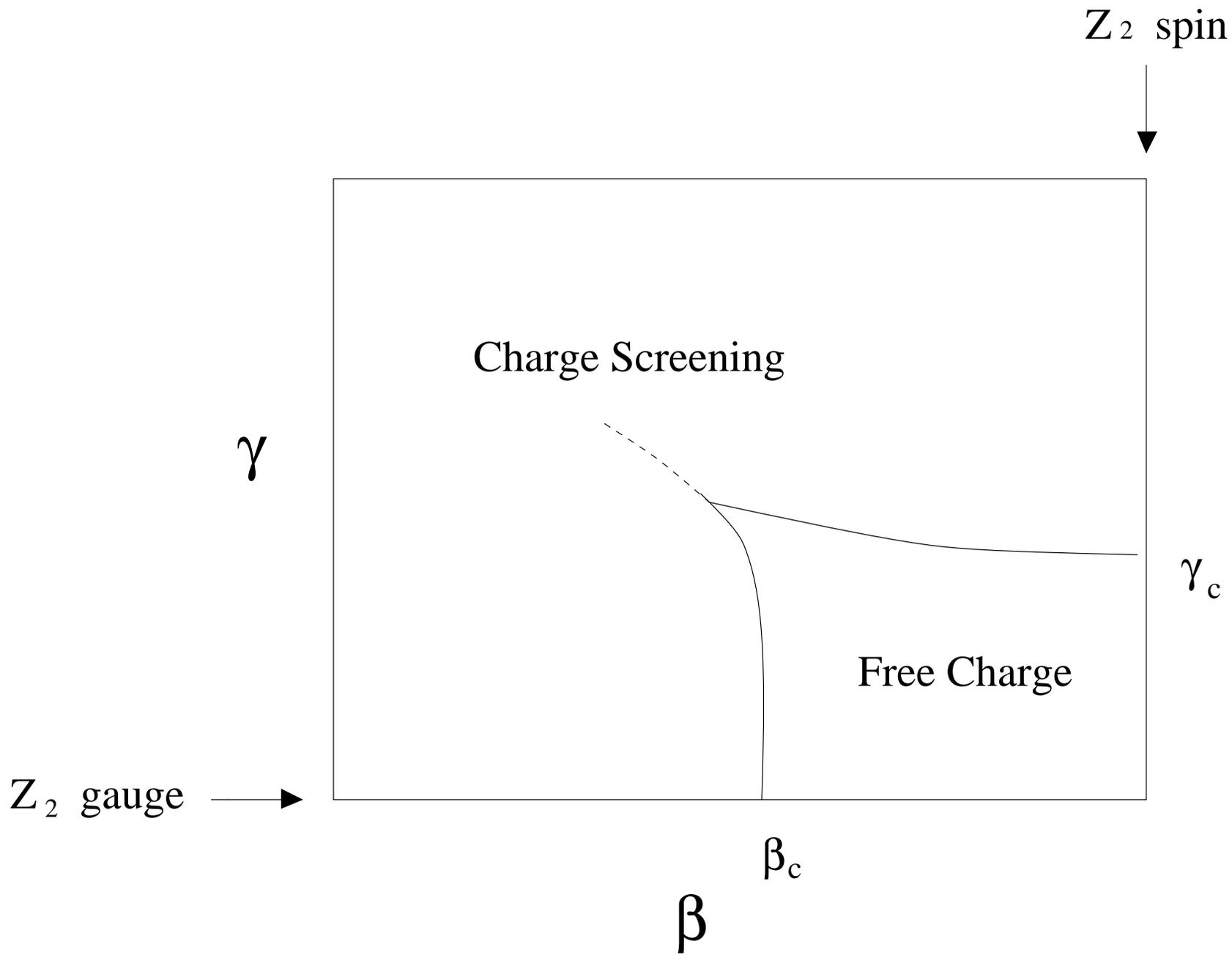
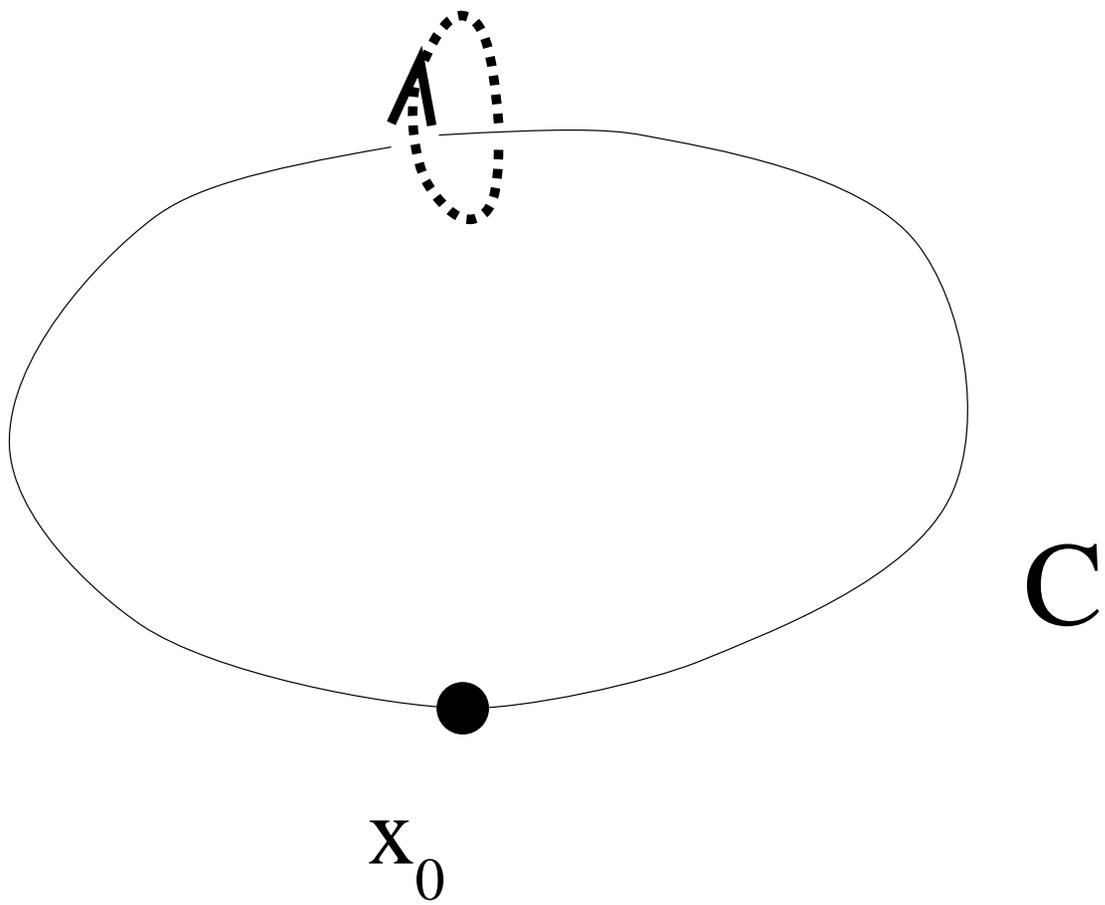
\end{document}